\RequirePackage[modulo]{lineno}
\documentclass[twocolumn,showpacs,aps,floatfix,prd,nofootinbib]{revtex4}
\usepackage{graphicx}
\usepackage{amsmath}
\usepackage[percent]{overpic}

\newcommand{\BABARPubYear}{17}
\newcommand{\BABARPubNumber}{003}
\newcommand{\SLACPubNumber}{17204}
\newcommand{\BaBarType}{PUB} 

\input babarsym							

\def\Ecm       {\ensuremath {E_{\rm c.m.}}\xspace}
\def\jetset   {\mbox{\tt Jetset \hspace{-0.5em}7.\hspace{-0.2em}4}\xspace}
\def\jetsett   {\mbox{\tt Jetset }\xspace}

\long\def\inst#1{\par\nobreak\kern 4pt\nobreak
    {\it #1}\par\vskip 10pt plus 3pt minus 3pt}

\begin{document}
\begin{flushleft}
\babar-\BaBarType-\BABARPubYear/\BABARPubNumber \\
SLAC-PUB-\SLACPubNumber \\
\end{flushleft}

\title{\large \bf
\boldmath
Study of the process $\epem \to \pipi\eta $ using initial state 
radiation 
}

%
\author{J.~P.~Lees}
\author{V.~Poireau}
\author{V.~Tisserand}
\affiliation{Laboratoire d'Annecy-le-Vieux de Physique des Particules (LAPP), Universit\'e de Savoie, CNRS/IN2P3,  F-74941 Annecy-Le-Vieux, France}
\author{E.~Grauges}
\affiliation{Universitat de Barcelona, Facultat de Fisica, Departament ECM, E-08028 Barcelona, Spain }
\author{A.~Palano}
\affiliation{INFN Sezione di Bari and Dipartimento di Fisica, Universit\`a di Bari, I-70126 Bari, Italy }
\author{G.~Eigen}
\affiliation{University of Bergen, Institute of Physics, N-5007 Bergen, Norway }
\author{D.~N.~Brown}
\author{Yu.~G.~Kolomensky}
\affiliation{Lawrence Berkeley National Laboratory and University of California, Berkeley, California 94720, USA }
\author{M.~Fritsch}
\author{H.~Koch}
\author{T.~Schroeder}
\affiliation{Ruhr Universit\"at Bochum, Institut f\"ur Experimentalphysik 1, D-44780 Bochum, Germany }
% ...
\collaboration{The \babar\ Collaboration}
\noaffiliation

\begin{abstract}
We study the process $\epem \to \pipi\eta \gamma$, where the photon is radiated from the initial state.
About 8000 fully reconstructed events of this process are selected from the \babar\ data sample with an integrated luminosity of 469~\invfb. Using the $\pipi\eta$ invariant mass spectrum we measure the  
$\epem \to \pipi\eta$ cross section in the \epem center-of-mass energy range from 1.15 to 3.5 \gev. The cross section is well described by the Vector-Meson Dominance model with four $\rho$-like states. We observe $49\pm9$ events of the $J/\psi$ decay to $\pipi\eta$, and measure the product $\Gamma_{J/\Psi \to \epem} \BR_{J/\Psi \to \pipi\eta} = 2.34 \pm 0.43_{\text{stat}} \pm 0.16_{\text{syst}}$~\ev.
\end{abstract}
\maketitle
\setcounter{footnote}{0}

\section{\boldmath Introduction\label{intro}}
A photon radiated from the initial state in the reaction $\epem \to \gamma + \text{hadrons}$ 
effectively reduces the electron-positron collision energy. This allows the study of hadron 
production over a wide range of \epem center-of-mass energies in a single experiment. The 
possibility of exploiting initial-state-radiation (ISR) events to measure low-energy cross 
sections at high-luminosity $B$ factories is discussed in 
Refs.~\cite{NLO_ISR,ISRprinciple,ISRprinciple1} and motivates the study described in this 
paper. The study of ISR events at the $B$ factories provides independent cross section 
measurements and contributes to understanding low-mass hadron spectroscopy. 
\begin{figure}
\includegraphics[width=.4\textwidth]{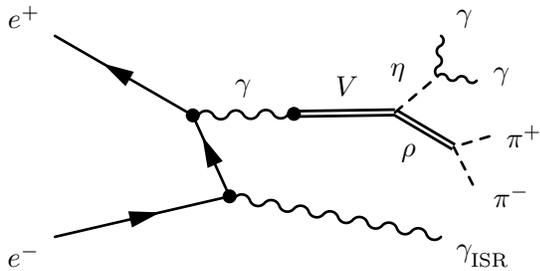}
\caption{The Feynman diagram for the process $\epem \to
\gamma_{\rm ISR}\pip\pim\gamma\gamma$ via the $\rho\eta$ intermediate state
in the Vector-Meson Dominance model.
\label{diagram}}
\end{figure}

In $\epem$ annihilations, final states like $\pip\pim\eta$ with positive G-parity must result from the isovector part of the hadronic current. Within the context of the Vector-Meson Dominance (VMD) model~\cite{NNAChasov}, the $\epem\to\pip\pim\eta$ process can be described by the Feynman diagram in Fig.~\ref{diagram}, where V represents any $\rho$ resonance, and $\rho$ is any accessible $\rho$ resonance.
The process is important for the determination of the parameters of $\rho$ resonances, gives a sizable contribution to the total hadronic cross section in the energy range 1.35--1.85 GeV. Additionally, results of the research can be used to test the relation between the $\epem \to \pipi\eta$ cross section and the spectral function for the decay $\tau^-\to\pi^-\piz\eta \nu_{\tau}$ predicted under the conserved
vector current (CVC) hypothesis~\cite{CVCEidelman}.

The process $\epem\to \pipi\eta$ was studied in several direct \epem experiments at energies from threshold to 2.4 GeV: DM1~\cite{DM1}, ND~\cite{ND}, DM2~\cite{DM2}, CMD-2~\cite{CMD-2}, and SND~\cite{2pieta_SND, 2pieta_SND2014}. This process was also studied by \babar\ using the decay mode $\eta \to \pi^- \pi^+ \pi^0$ with the ISR technique. The \babar\ study was based on a 239~\invfb data sample~\cite{BaBar2007}  and reached 3 GeV. The cross section    and \pip\pim mass distributions were consistent with VMD.  A theoretical study of the process $\epem\to \pipi\eta$ within VMD and Nambu-Jona-Lasinio chiral approaches was performed in Ref.~\cite{NNAChasov} and Refs.~\cite{NJL,ResonChirTh}, respectively.

This paper reports a study of the $\pipi\eta$ hadronic final state with $\eta\to 2\gamma$ 
produced together with a energetic photon that is assumed to result from ISR.
The invariant mass of the hadronic system determines the reduced effective \epem center-of-mass (c.m.) energy (\Ecm$\equiv m_{\pipi\eta}c^{2}$), and we measure the $\epem\to\pipi\eta$ cross section in the range $1.15 < \Ecm <3.5$~\gev. The different $\eta$ decay mode makes this independent of our previous work. We fit the results using the VMD model and extract $\rho$ resonances parameters, and we calculate a $\tau\to\pip\piz\eta\nu_\tau$ branching fraction under the CVC hypothesis.
   
\begin{figure*}
\includegraphics[width=.46\textwidth]{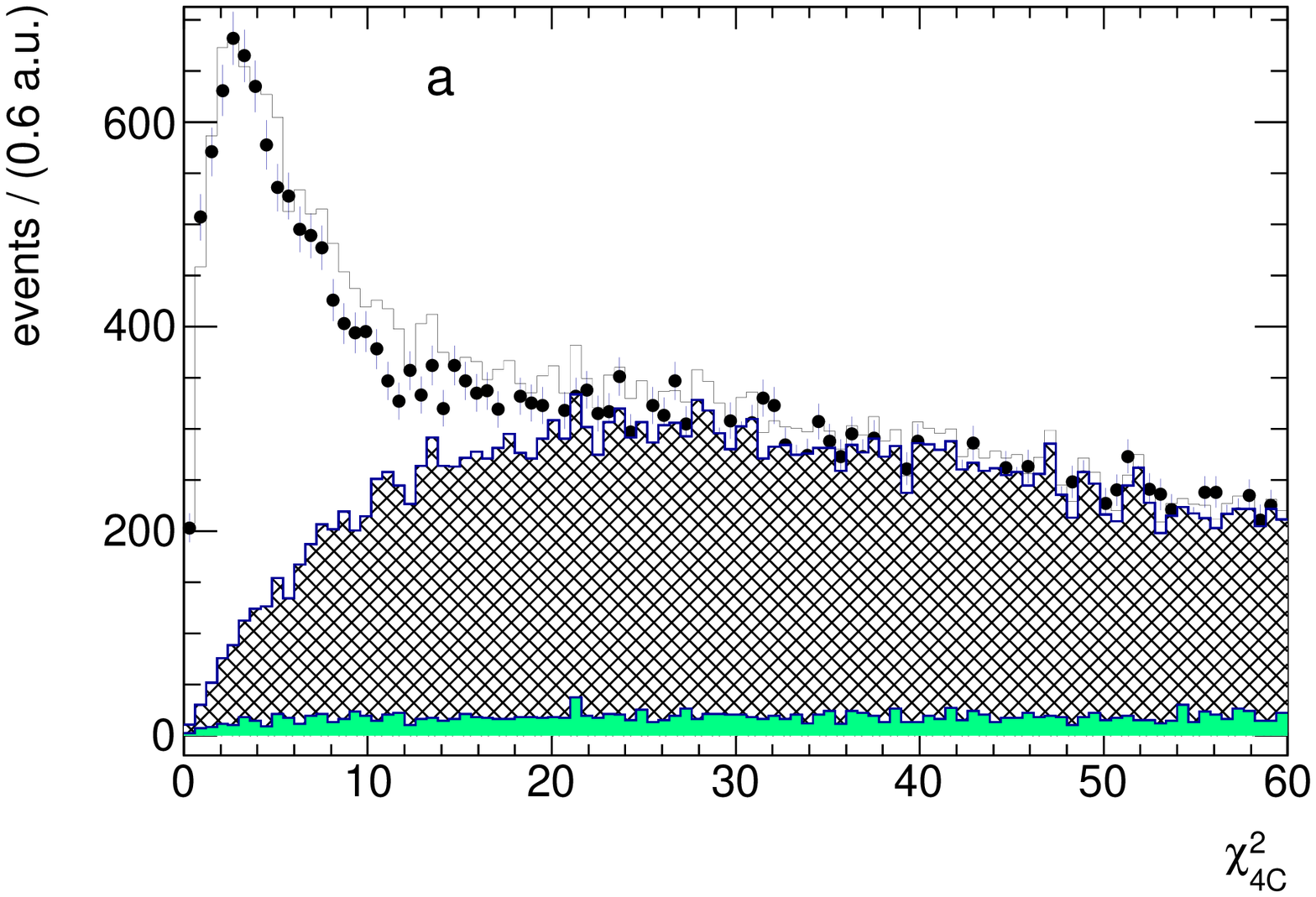} 
\includegraphics[width=.46\textwidth]{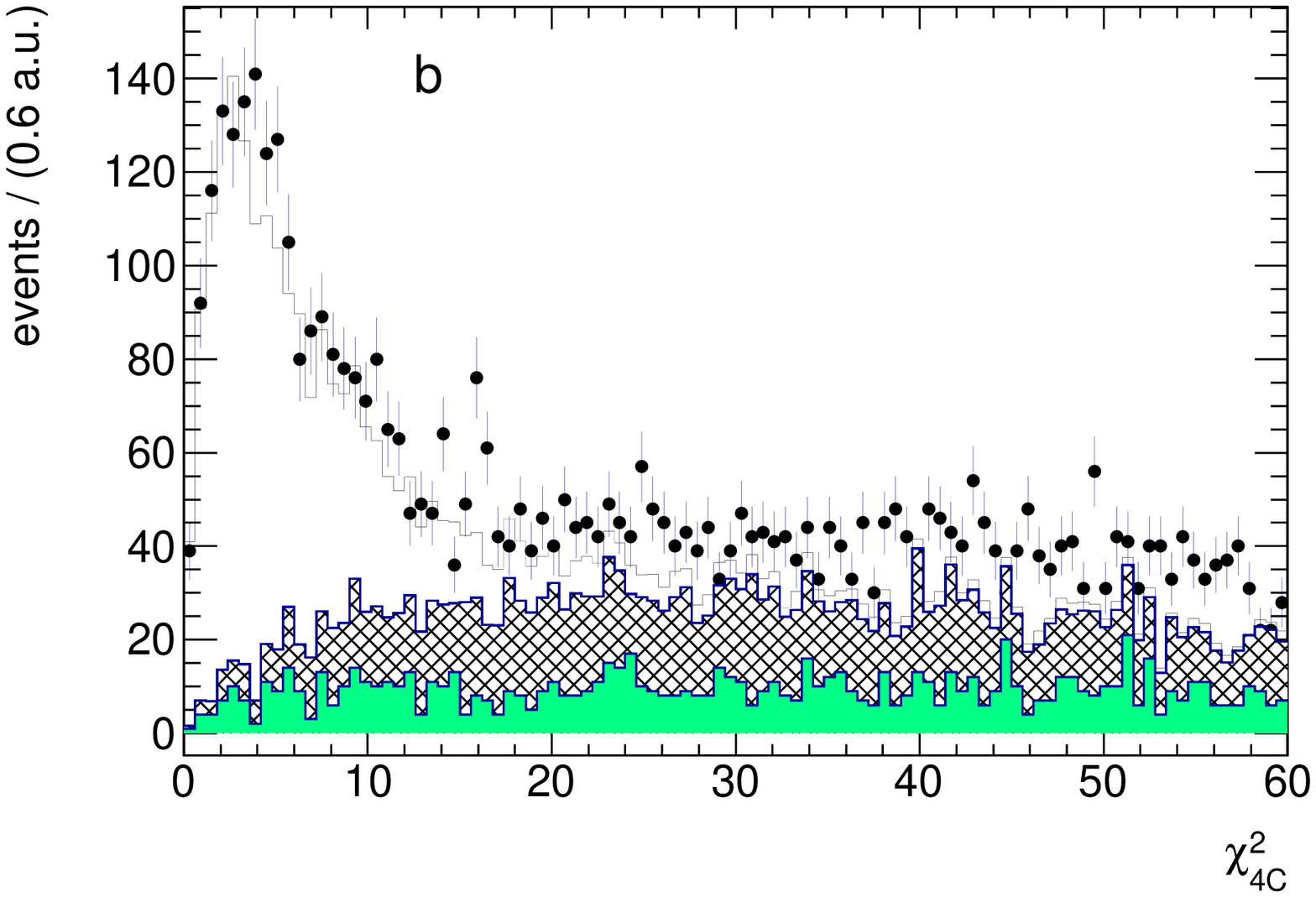} 
\caption{ (color online) The distributions of $\chi^2_{\rm 4C}$ for events from the invariant 
mass ranges $1.15 < m_{\pipi\eta} < 2.00$~\gevcc (a) 
and $2.0 < m_{\pipi\eta} < 3.5$~\gevcc (b). 
The points with error bars are data. The open histogram represents the sum of the simulated distributions for signal and backgrounds events. The shaded histogram represents non-ISR background, while the hatched 
area shows ISR background.}
\label{chi_1_3}
\end{figure*}

\section{\boldmath The \babar\ detector and data set}
\label{sec:babar}

The data used in this analysis were collected with the \babar\ detector at the
\pep2\ asymmetric-energy \epem\ collider at the SLAC National Accelerator 
Laboratory. The integrated luminosity of 468.6~\invfb~\cite{lumi} used in this analysis comprises 424.7~\invfb collected at the $\Upsilon(4S)$ resonance, 
and 43.9~\invfb collected 40 MeV below the peak.

The \babar\ detector is described in detail elsewhere~\cite{Detector, Detector1}. 
Charged particles are reconstructed using a tracking system, which comprises
a silicon vertex tracker (SVT) and a drift chamber (DCH) inside a 1.5 T 
solenoid magnet. Separation of pions and kaons is accomplished by means of the 
detector of internally reflected Cherenkov light (DIRC) and energy-loss measurements
in the SVT and DCH. The energetic ISR photon and photons from $\piz$ and $\eta$ 
decays are detected in the electromagnetic calorimeter (EMC).  
Muon identification is provided by the instrumented flux return of the magnetic field.

To study the detector acceptance and efficiency,   
a special package of programs for simulation of ISR processes was developed 
based on the approach suggested in Ref.~\cite{EVA}.  
Multiple collinear soft-photon emission from the initial \epem state 
is implemented with the structure-function technique~\cite{structuremethod}, 
while additional photon radiation from the final-state particles (FSR) is
simulated using the PHOTOS package~\cite{photos}.  
The precision of the radiative-correction simulation does not contribute more
than 1\% uncertainty to the efficiency calculation.

The process $\epem\to \pipi\eta\gamma$ is simulated assuming the intermediate $\rho(770)\eta$ hadronic state. Generated events are processed through the detector response simulation~\cite{GEANT4} and then reconstructed using the same procedure as the real data. Variations in the detector and background conditions are taken into account in the simulation.

We simulate the background ISR processes $\epem\to \Kp\Km\eta\gamma$, $\pipi \ppz\gamma$, $\pipi 3\piz\gamma$, $\pipi\piz\eta\gamma$, and $\pipi\piz\gamma$, and non-ISR processes $\epem\to\tau^+\tau^-$ and  $\epem\to q \qbar$ $(q = u, d, s)$. The latter process is generated using the \jetset~\cite{udssim} event generator.

\section{\boldmath Event selection and kinematic fit}
\label{event_select}

Preliminary selection criteria require detection of a high-energy photon with 
a c.m.\ energy greater than 3~\gev, at least two charged-particle tracks, and at least two additional photons with invariant mass near the $\eta$ mass, in the range 0.44--0.64~\gevcc. 
Each of the photons is required to have an energy greater than 100~\mev\footnote{Unless otherwise specified, all quantities are evaluated in the laboratory frame} and a polar angle in the range 0.3--2.1 radians. The photon with the highest c.m. energy is 
assumed to be from ISR. Charged-particle tracks are required to originate within 0.25 
cm of the beam axis and within 3 cm of the nominal collision point along the 
axis. Each of the tracks is required to have momentum higher than 100~\mevc,
and be in the polar angle range 0.4--2.4 radians. Additionally, the tracks 
are required to be not identified as kaons or muons.
If there are three or more tracks, the oppositely charged pair with closest
 distance to the interaction region is used for the further analysis. 
The selected  candidate events are subjected to a 4C kinematic fit under 
the $\epem\to\pipi 3\gamma$ hypothesis, which includes four constraints of 
energy-momentum balance. The common vertex of the charged-particle 
tracks is used as the point of origin for the detected photons.
There is no constraint on the $\eta$ candidate mass, since this will 
be used below to extract the number of signal events.
Monte-Carlo (MC) simulation and data samples contain a significant number of 
false photons arising from split-off charged-pion EMC clusters and 
beam-generated background, as well as additional ISR or FSR photons. 
For events with more than three photons we perform a kinematic
fit for all photon-pair combinations not including the ISR photon, and choose the combination with the 
lowest value of $\chi^2_{\rm 4C}$.  The parameter $\chi^2_{\rm 4C}$ is 
used to discriminate between signal and background events.

Since the production of the two-pion system is predominantly via $\rho$-meson intermediate states we require that the invariant mass of the two pions, $m_{2\pi}$, is greater than 0.4 GeV/c$^2$. Because of very different background conditions, the $\pipi\eta$ invariant 
mass interval under study is divided into two regions: 
$1.15 < m_{\pipi\eta} < 2.00$~\gevcc (I) and 
$2.0 < m_{\pipi\eta} < 3.5$~\gevcc (II).
Two additional selection conditions are used for Region II: the energies of 
photons from the $\eta$ decay are required to be greater than 200 MeV and 
$m_{\pi^{\pm}\gamma_{\text{ISR}}} > 1$ GeV/c$^{2}$, where $m_{\pi^{\pm}\gamma_{\text{ISR}}}$ is 
the invariant mass of the charged pion and the ISR photon. 
The latter condition rejects 
$e^{+}e^{-} \to \tau^{+} \tau^{-}$ background events with one of the $\tau$ decaying into $\rho^\pm\nu\to\pi^\pm\pi^0\nu$, where
an energetic photon, considered as $\gamma_{\rm ISR}$, arises from $\pi^0$ decay. In this case
 the spectrum of invariant mass of the most energetic photon and one of 
the selected charged pions is peaked near the $\rho$ mass.
\begin{figure*}
\includegraphics[width=.4\textwidth]{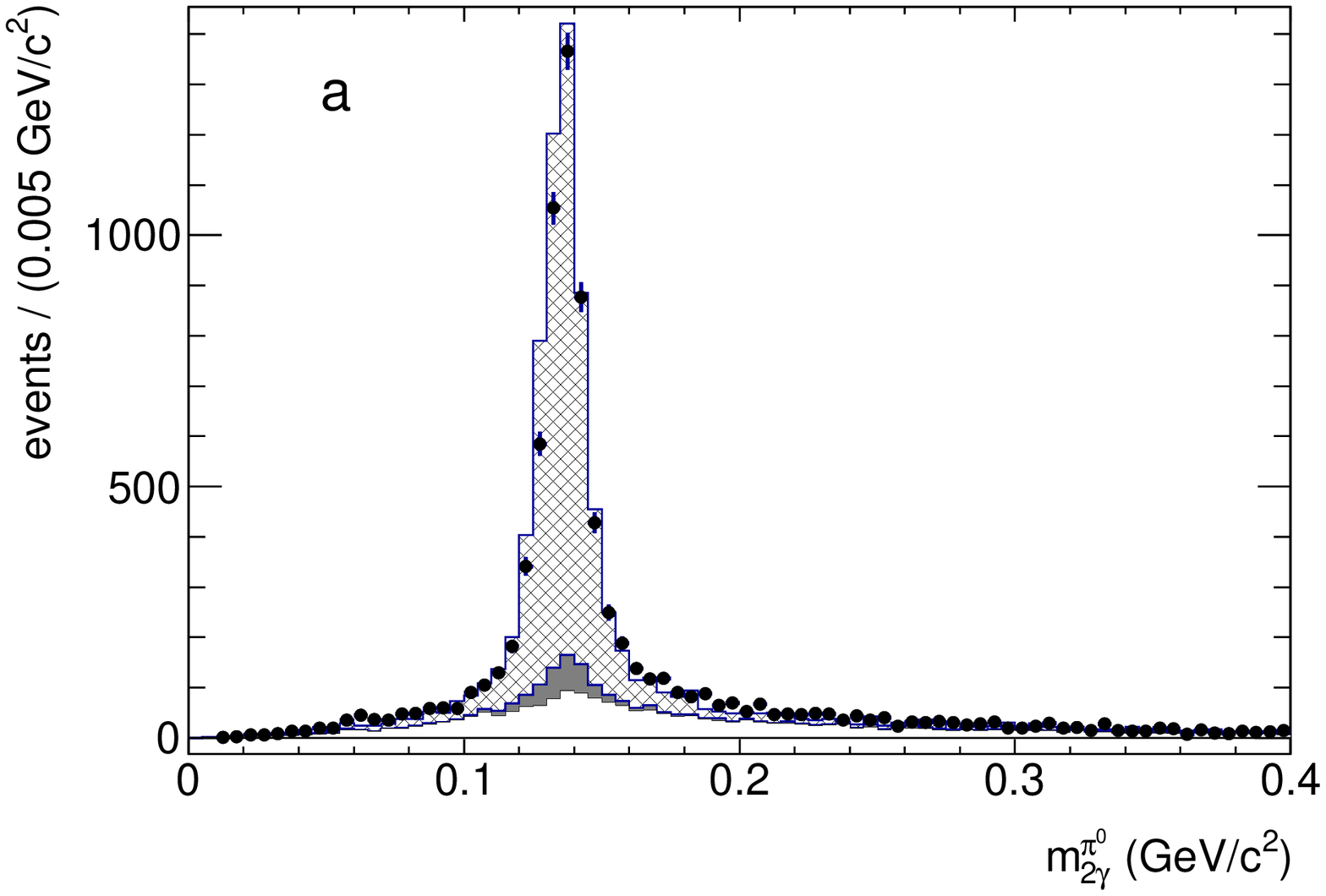}
\includegraphics[width=.4\textwidth]{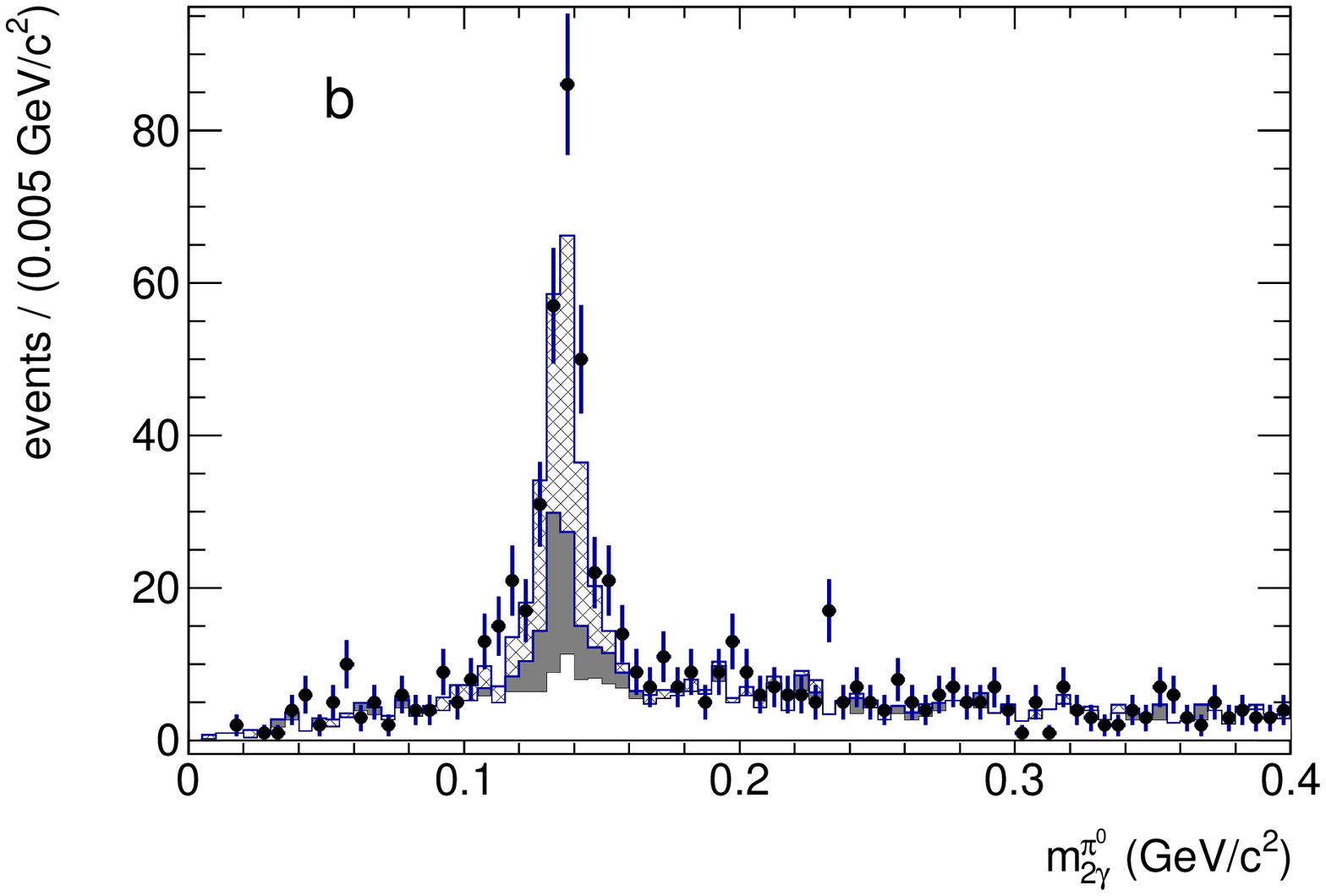}
\caption{The distribution of $m^{\pi^{0}}_{2\gamma}$ for Region I (a) and 
for Region II (b). 
The points with error bars are data. The open histogram represents signal 
simulation, the shaded and hatched areas shows simulated non-ISR and ISR 
background contributions, respectively.}
\label{m2phrest}
\end{figure*}
  
The $\chi^2_{\rm 4C}$ distributions for events from Region I and Region II
are shown in Fig.~\ref{chi_1_3}. The points with error bars represent data, 
while the histograms show, cumulatively, the contributions of simulated 
non-ISR background (shaded), ISR background (hatched), and signal 
$\epem\to\pipi\eta\gamma$ events (open histogram). For background, 
the distributions are normalized to the expected numbers of events calculated 
using known experimental cross sections, in particular, 
\cite{2pi2pi0datababar} for $\epem\to \pipi \ppz \gamma$, 
\cite{babar2keta} for $\epem\to \Kp\Km\eta \gamma$,
 \cite{snd3pieta} for  $\epem\to\pipi\piz\eta\gamma$ and \cite{tautausim} for $e^+e^-\to \tau^+\tau^-$.
For the non-ISR $\epem\to q \qbar$ background, the expected number is corrected
to take into account the observed data-MC simulation difference (see below).
The signal distribution is normalized in such a way that the total 
simulated distribution matches the first seven bins of the data distribution.
It is seen that the simulated backgrounds at $\chi^2_{\rm 4C} > 20$ are adequate in the lower-mass $m_{\pipi\eta}$ region, but not in the higher.
The conditions $\chi^2_{\rm 4C}< 25$ and $\chi^2_{\rm 4C} < 15$ are used
for Region I and II, respectively.

Most background processes contain neutral pions in the final state. To 
suppress this background, we check all possible combinations of pairs of photons 
with energy higher than 100 MeV and choose the one with invariant mass 
($m^{\pi^{0}}_{2\gamma}$) closest to the $\pi^{0}$ mass. The obtained 
$m^{\pi^{0}}_{2\gamma}$ distribution is shown in Fig.~\ref{m2phrest}. 
We apply the requirement $m^{\pi^{0}}_{2\gamma} > 0.16$ GeV/c$^{2}$.
With these conditions, 11469 data events are selected.

The remaining simulated ISR background is still dominated by the 
$\epem\to\pipi\piz\piz\gamma$ process. In the non-ISR background, about 50\%
of events come from the process $\epem\to q \qbar\to\pipi\piz\eta$,
which imitates the process under study when one of photons from the \piz decay
is soft and the other is identified as the ISR photon. 
Such events preferentially have a small $\chi^2_{\rm 4C}$ like signal events. 
Remaining non-ISR events come from the process $\epem\to q \qbar\to\pi^-\pi^+\pi^0\pi^0$
or from processes with higher neutral particle multiplicity 
($\epem\to q \qbar\to\pi^{+}\pi^-\pi^0\pi^0\pi^0$, 
$\epem\to q \qbar\to\pi^-\pi^+\eta\pi^0\pi^0$, etc.), and have a uniform $\chi^2_{\rm 4C}$ 
distribution. 
To check the quality of the \jetsett simulation, we select non-ISR events in data 
and simulation using the following procedure. 
\begin{figure}
\includegraphics[width=.4\textwidth]{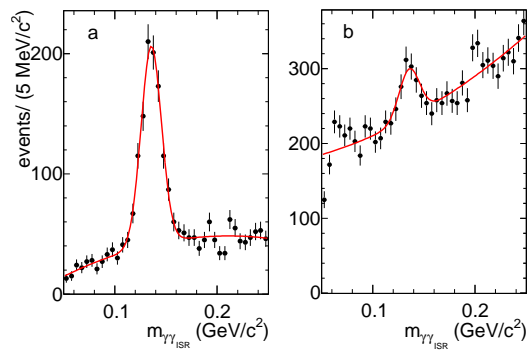} 
\caption{The distributions of the invariant mass
for all combinations of the ISR-photon candidate with any other photon in an event in 
$\epem\to q\bar{q}$ simulation (a) and data (b). The curves are the
results of the fits described in the text.
\label{uds}}
\end{figure}
We remove the condition $m^{\pi^{0}}_{2\gamma} > 0.16$ GeV/c$^{2}$ and modify the
$\chi^2_{\rm 4C}$ condition to $\chi^2_{\rm 4C} < 100$. The invariant masses
for all combinations of the ISR-photon candidate with any other photon in an event are calculated. The mass distributions are shown in  Fig.~\ref{uds} for simulated $q\bar{q}$ and data events. The \piz peak is clearly seen both in data and in simulation, indicating the presence of non-ISR processes. The distributions are fitted with a sum of a Gaussian function describing the \piz resolution function and a second-order polynomial. In the fit to the data distribution, the parameters of the Gaussian function are fixed to the values obtained in the fit to the simulated distribution. The ratio of the number of data events in the \piz peak to that expected from the \jetsett simulation is found to be $0.70\pm0.05$. This data-MC simulation scale factor is an average over the mass range $1.15 < m_{\pipi\eta} < 3.5$~\gevcc. We do not observe a $m_{\pipi\eta}$  dependence of the scale factor at the level of the available statistics.  After the simulation normalization the number of $\epem\to q \qbar\to\pipi\piz\eta$ events satisfying our standard selection criteria is estimated to be 171$\pm$12.
\begin{figure*}
\begin{overpic}[scale=0.4]{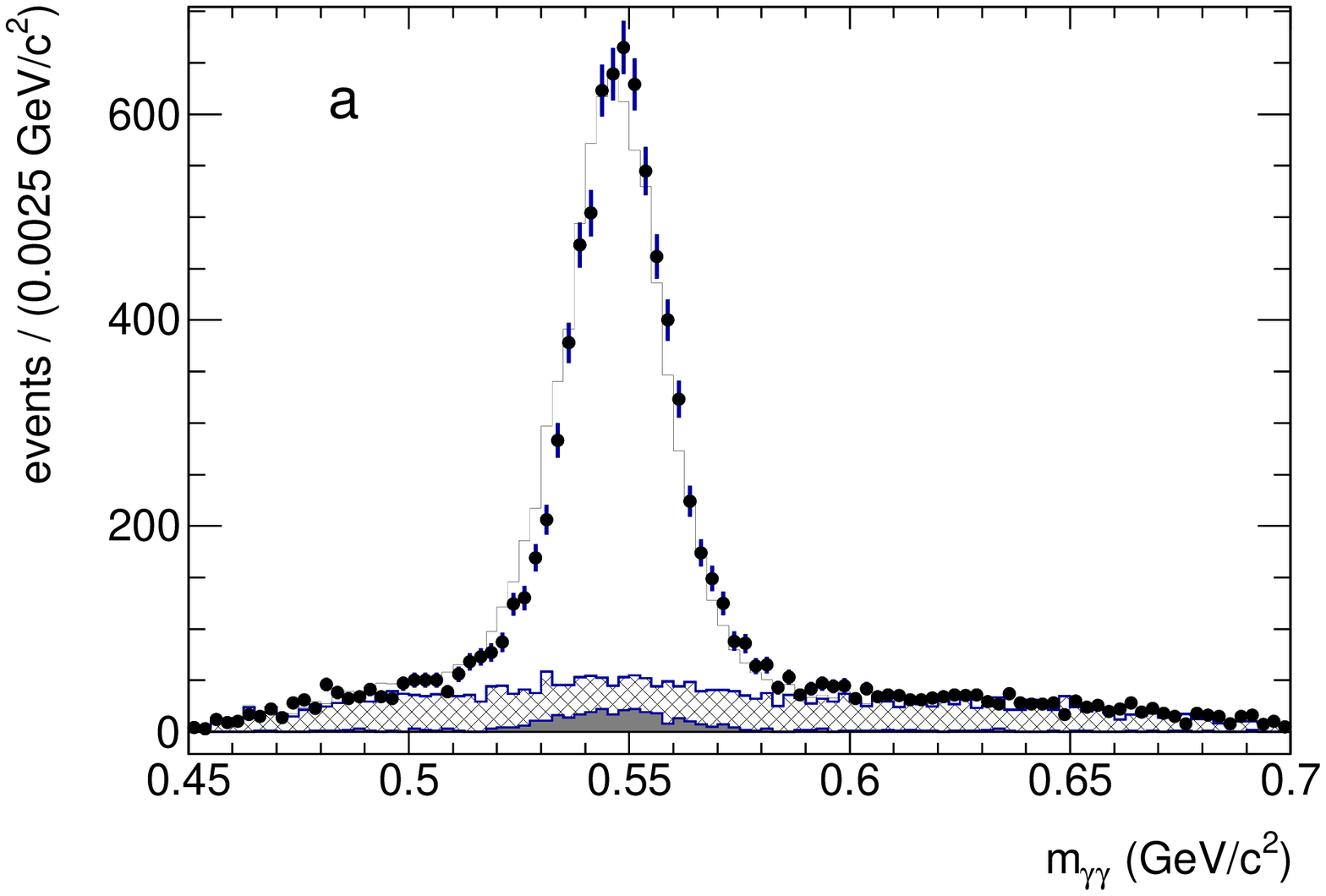} 
\end{overpic}
\includegraphics[width=80mm]{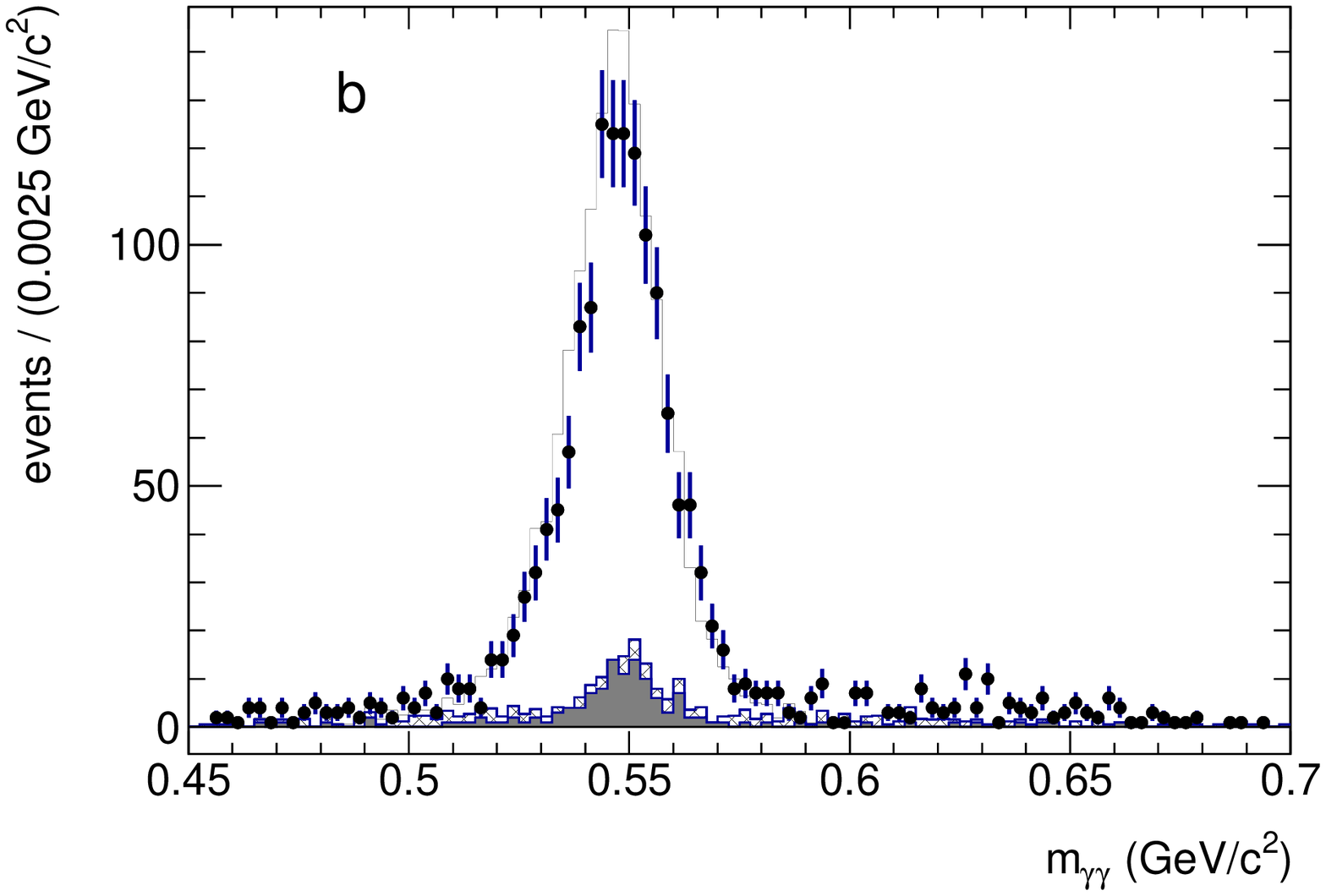}
\caption{The two-photon invariant mass distribution for events from Region I 
(a), and Region II (b) after all selections. The points with error bars 
represent data. The histograms cumulatively show simulated contributions for 
peaking background (shaded), nonpeaking background (hatched), and signal 
$\epem\to \pipi\eta\gamma$ events (open). 
\label{chichi2}}
\end{figure*}
\begin{figure}
\includegraphics[width=85mm]{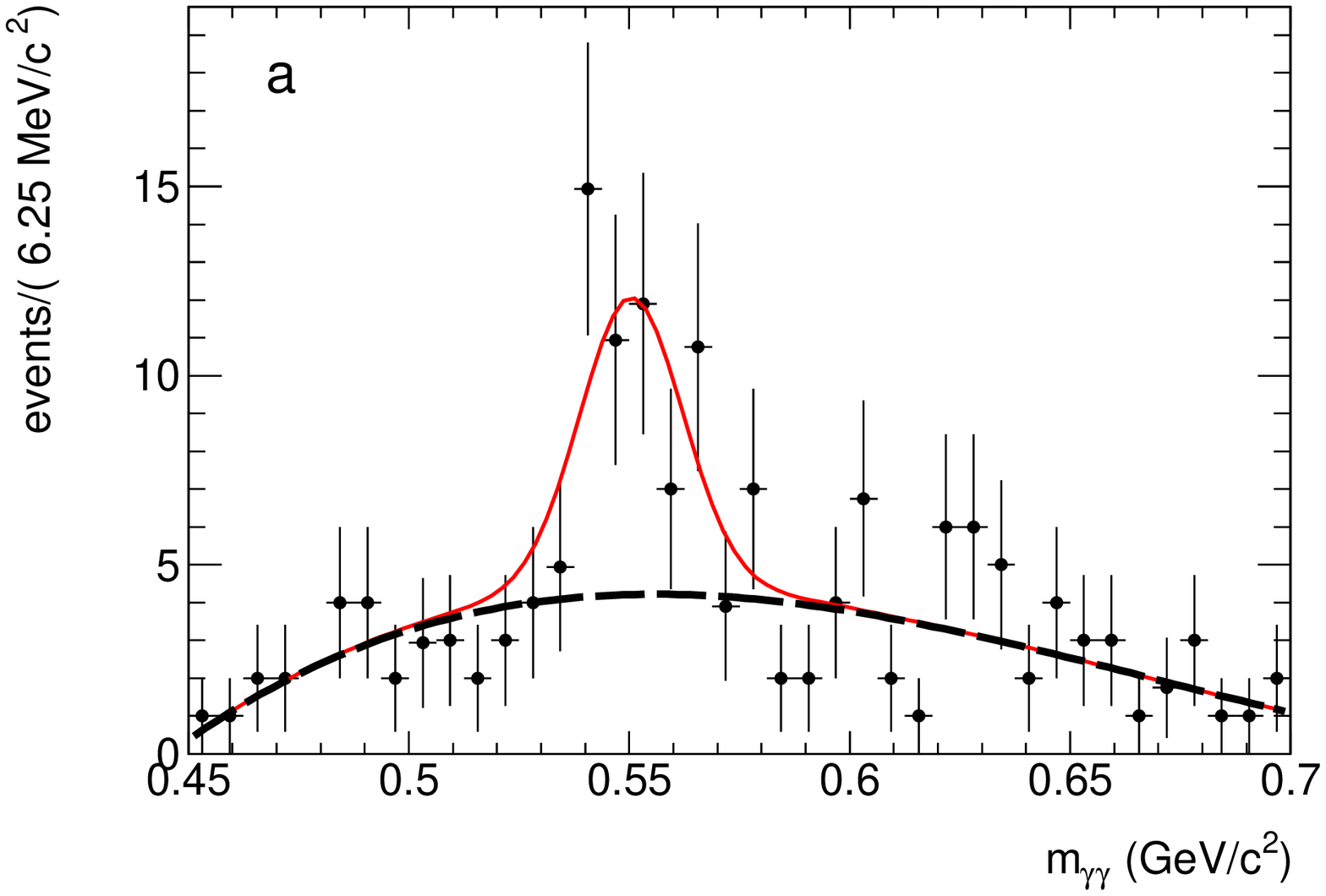}
\includegraphics[width=85mm]{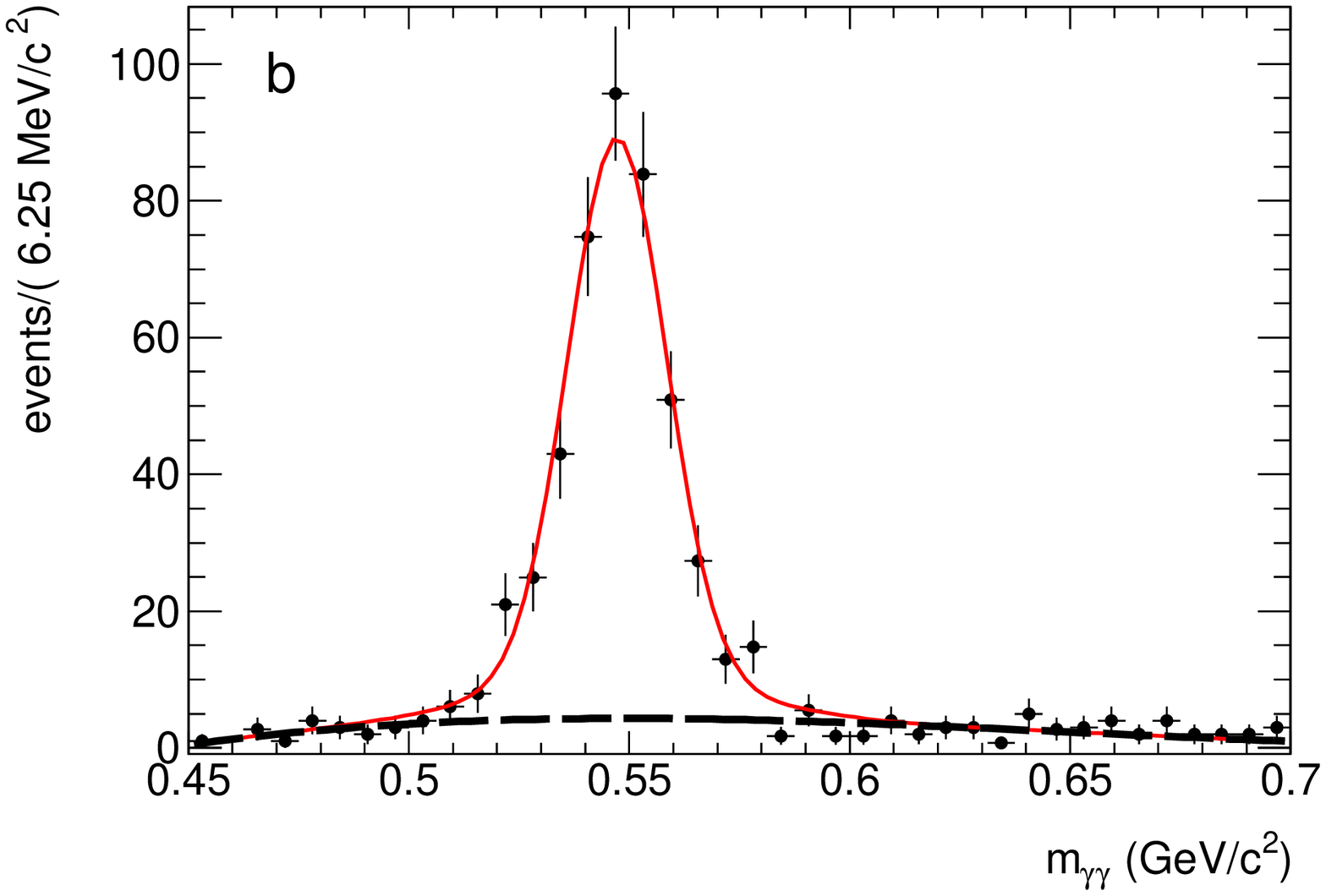}
\includegraphics[width=85mm]{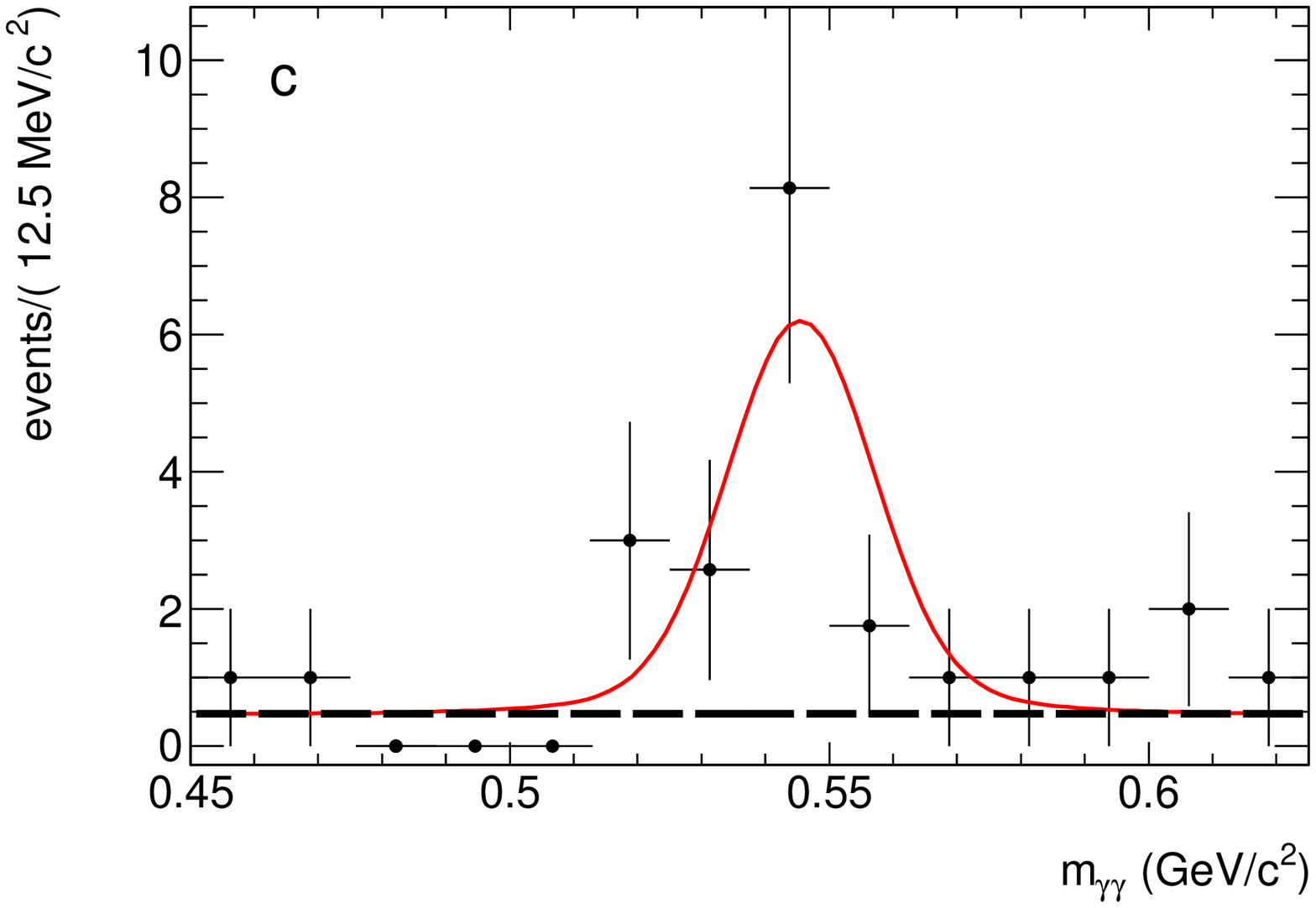}
\caption{The two-photon invariant mass spectrum for data events (points with
error bars) from the three $m_{\pipi\eta}$ 
intervals: $1.300-1.325$ ~\gevcc (a), $1.500-1.525$~\gevcc (b), and
$3.4-3.5$~\gevcc (c). The solid curve is the result of the fit described in
the text. The dashed curve represents the fitted background.
\label{mgg_sub}}
\end{figure}

\section{Background subtraction\label{signcalculation}}
Figure~\ref{chichi2} shows the distribution of the $\eta$-meson candidate 
invariant mass ($m_{\gamma\gamma}$) for Regions I and II. The invariant
mass is calculated using the photon parameters returned by the 4C kinematic
fit. The points with error bars represent data. The open histograms show the 
$m_{\gamma\gamma}$ distribution for signal simulated events. The shaded and hatched
histograms show the expected contributions from background events peaking and
nonpeaking at the $\eta$-meson mass, respectively.
The peaking background arises from the processes 
$\epem\to\pipi\piz\eta$, $\epem\to K^{+}K^{-}\eta\gamma$, and 
$\epem\to \pi^{+}\pi^{-}\pi^{0}\eta\gamma$.
\begin{figure*}
\begin{overpic}[scale=0.5]{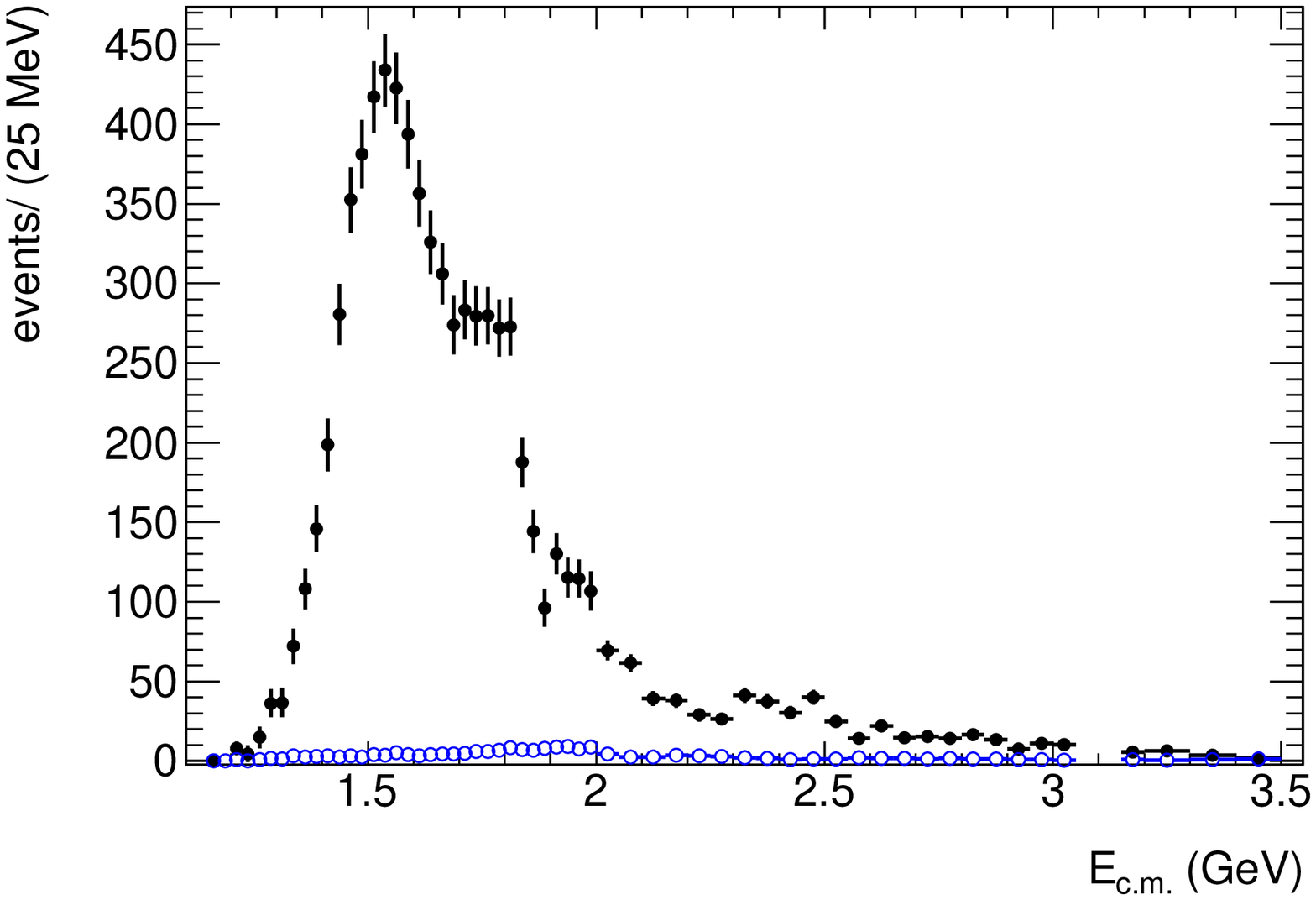}
\put(44,29){\includegraphics[scale=0.24]{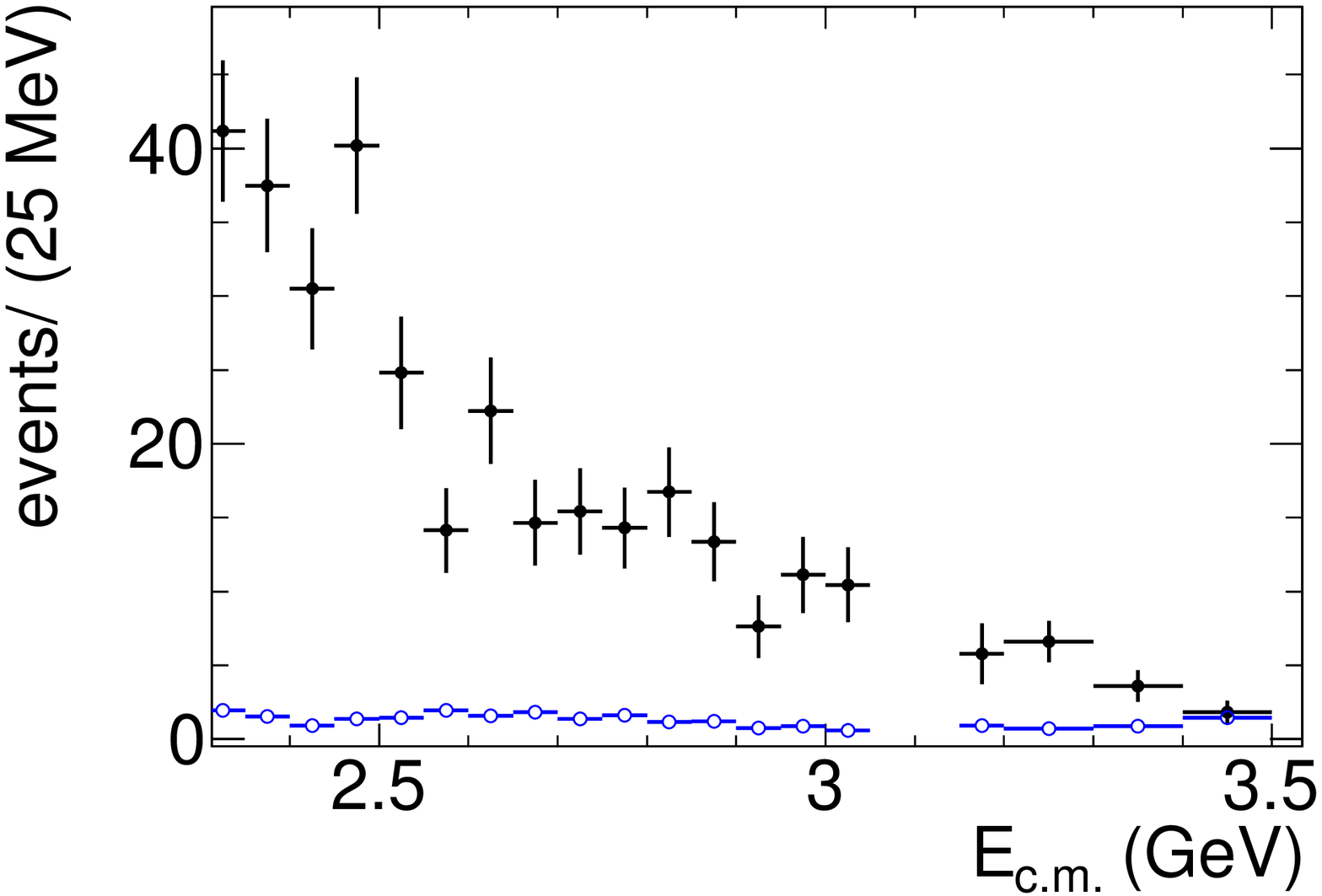}}
\end{overpic}
\caption{The measured $\pipi\eta$ invariant mass spectrum (solid circles).
The open circles show the same distribution for the peaking background 
obtained using MC simulation.}
\label{Nevents}
\end{figure*}

The number of signal events is determined from the fit to the $m_{\gamma\gamma}$ 
spectrum by a sum of signal and background distributions. The signal line shape is 
described by a double-Gaussian function, the parameters of which are obtained 
from MC simulation. The shape and the number of events for peaking background 
are calculated using MC simulation. In Region I, where simulation reproduces the
$m_{\gamma\gamma}$ spectrum reasonably well (see Fig.~\ref{chichi2}a), the 
nonpeaking background shape is taken from MC simulation. In 
Region II (Fig.~\ref{chichi2}b), the background shape is assumed to be 
uniform in the $m_{\gamma\gamma}$ range from 0.45 to 0.65~\gevcc. The free fit 
parameters are the numbers of signal events and number of nonpeaking
background events.  

The fit is performed in the 59 $m_{\pipi\eta}$ bins listed in 
Table~\ref{cross_table0}. The mass bin width is chosen to be 
25~$\rm MeV/$$c^2$ below 2.0~\gevcc, and 50 (100)~$\rm MeV/$$c^2$ in the 
range $2.0 < m_{\pipi\eta} < 3.1$ ($3.1 < m_{\pipi\eta} < 3.5$)~\gevcc.
Our measurement is restricted to the mass range
$1.15 < m_{\pipi\eta} < 3.50$~\gevcc.
Outside this range the signal to background ratio is too small to observe 
the signal. 
The fit results are shown in Fig.~\ref{mgg_sub} for three representative 
$m_{\pipi\eta}$ bins. The fitted number of signal events as a function of the 
$\pipi\eta$ invariant mass is shown in Fig.~\ref{Nevents} together with the 
$m_{\pipi\eta}$ spectrum for peaking background calculated using MC simulation.
The total number of signal events is
found to be $8065\pm101$, while the numbers of peaking and nonpeaking 
background events are  $239\pm18$ and $3164\pm64$, respectively.
\begin{figure*}
\begin{minipage}[t]{0.45\textwidth}
\includegraphics[width=.95\textwidth]{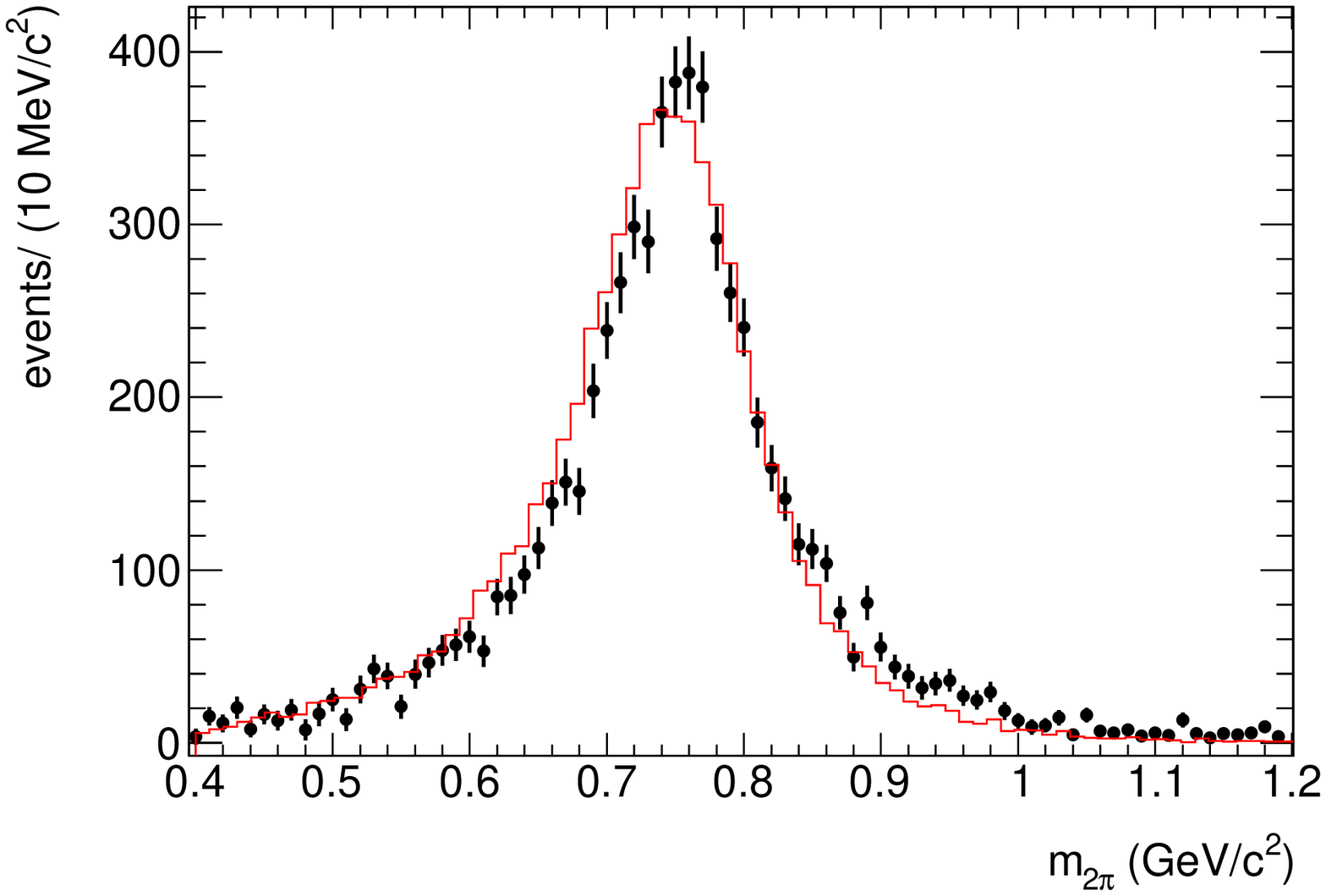}
\caption{The two-pion invariant mass distribution for data (points with error bars) and 
simulated (histogram) events from the mass range 
$1.4 < m_{\pipi\eta} < 2.0$~\gevcc.\label{m2pi}}
\end{minipage}
\hfill
\begin{minipage}[t]{0.45\textwidth}
\includegraphics[width=.95\textwidth]{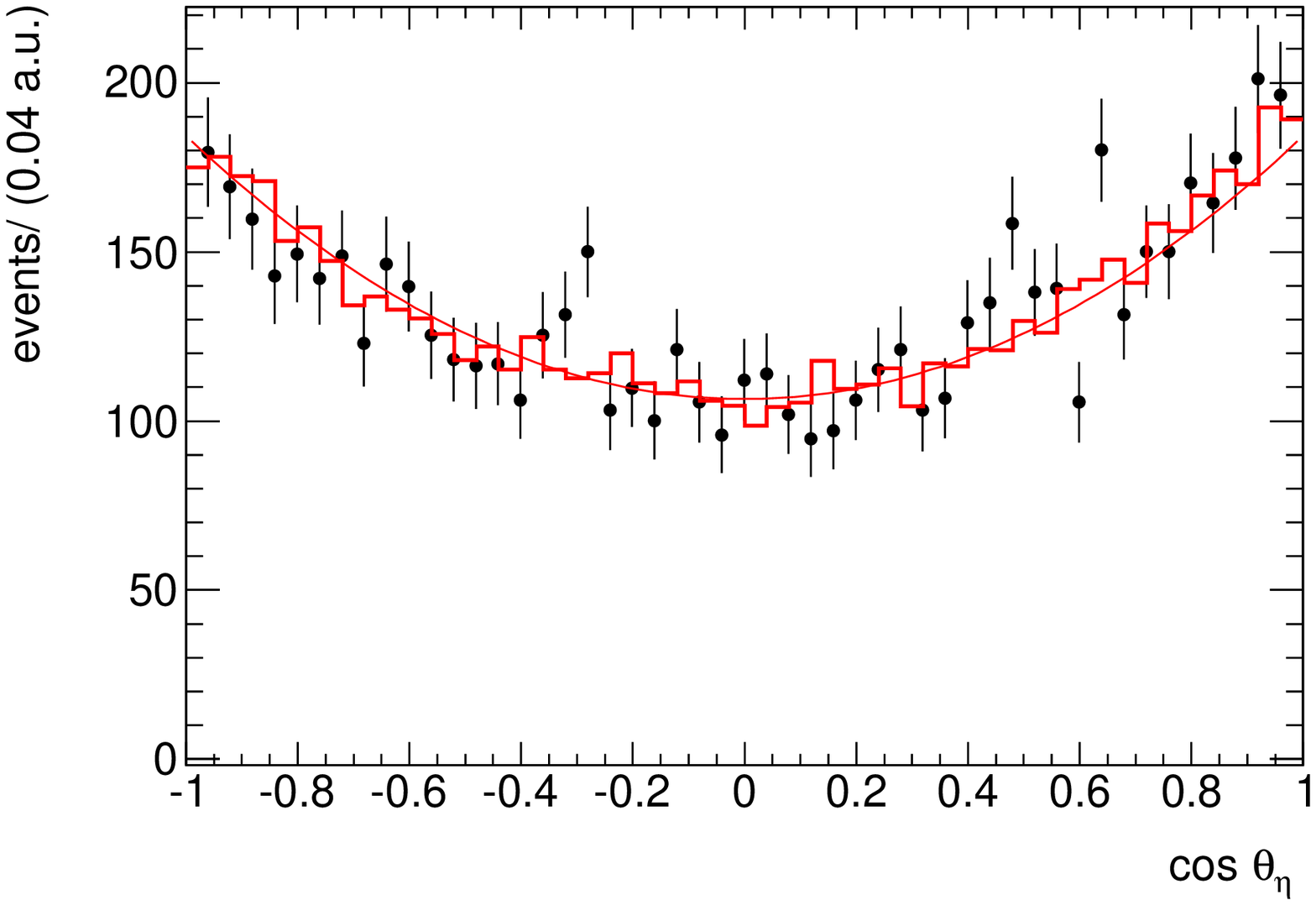}
\caption{The $\cos{\theta_{\eta}}$ distribution for data (points with error bars) and 
simulated (histogram) events from the mass range 
$1.4 < m_{\pipi\eta} < 2.0$~\gevcc. The curve is the result of the fit
described in the text.
\label{etaangle}}
\end{minipage}
\end{figure*}

A similar procedure of background subtraction is used to obtain the $\pipi$ 
invariant mass  spectrum for data events in the range 
$1.4 < m_{\pipi\eta} < 2.0$~\gevcc. The spectrum is shown in Fig.~\ref{m2pi}
in comparison with the simulated signal spectrum. The simulation uses the model
of the $\eta\rho(770)$ intermediate state.
The observed difference between data and simulated spectra 
may be explained by the contribution of other intermediate 
states, for example $\eta\rho$(1450), and their interference with the dominant
$\eta\rho(770)$ amplitude. This effect was observed previously in the SND 
experiment~\cite{2pieta_SND2014}. 

Figure~\ref{etaangle} shows the $\cos{\theta_\eta}$ distribution, where
$\theta_\eta$ is the angle between the $\eta$ momentum
in the $\pipi\eta$ rest frame and the ISR photon direction in the c.m. frame.
In the $\eta\rho$ model this distribution is expected to be $(1+\cos^{2}{\theta_{\eta}})$. However the detection efficiency of the process under study depends on cos $\theta_{\eta}$ and data events are distributed as $(1+(0.73\pm0.08)\cdot\cos^{2}{\theta_{\eta}})$ according to the fit shown by a curve in the figure. The detection efficiency is correctly reproduced in MC simulation and the distribution of reconstructed simulated events shown by a histogram in the figure is in reasonable agreement with data.

\begin{figure}
\includegraphics[width=90mm]{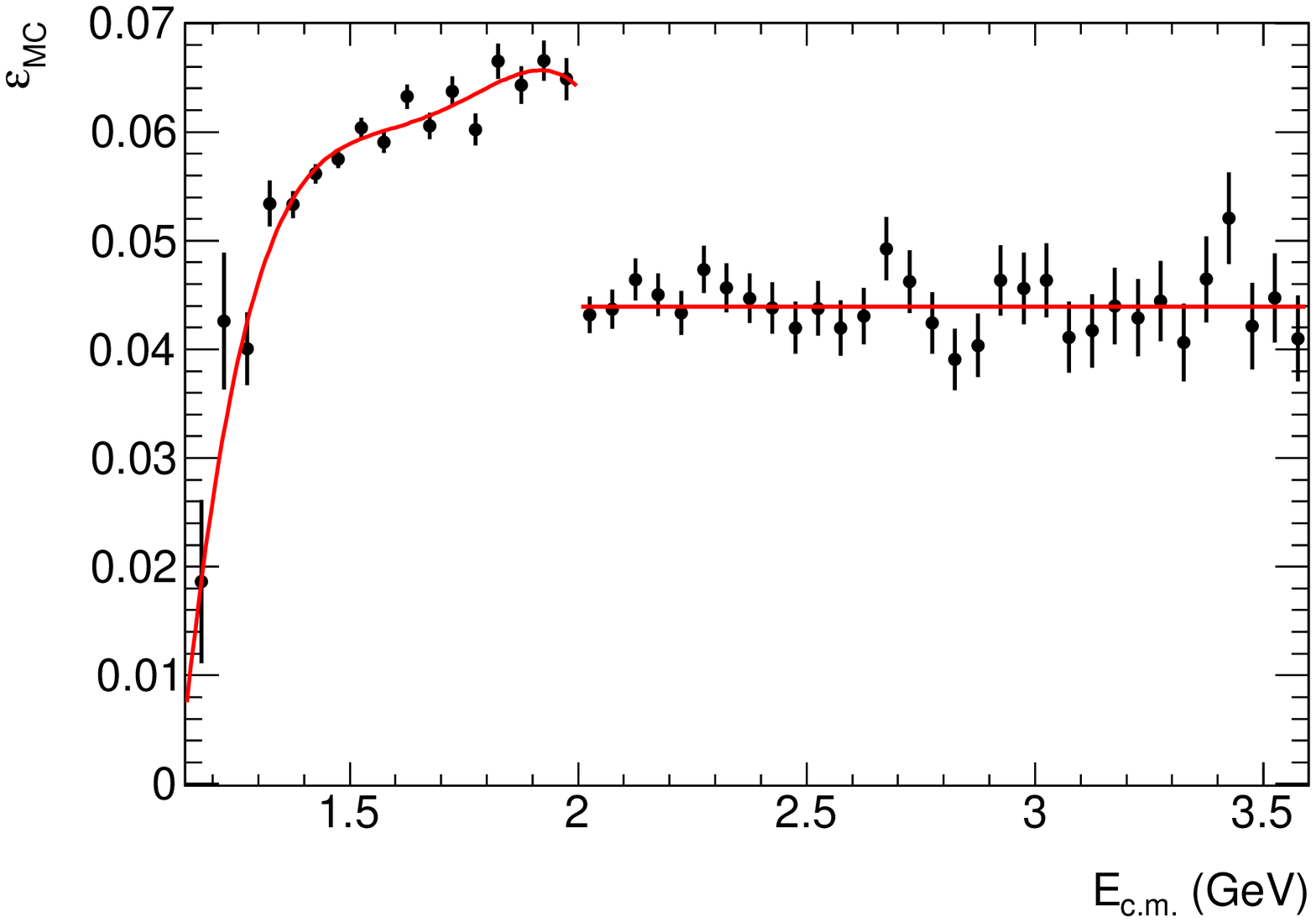}
\caption{The detection efficiency $\varepsilon_{\text{MC}}$ obtained using MC 
simulation as a function of	 $\Ecm$. The lines are fits to a fourth-order polynomial for $E_{c.m.} < 2$ GeV and to a constant for $E_{c.m.} > 2$ GeV.
\label{effmc}}
\end{figure}
\begin{table*}
\caption{Results of the $\epem \to \pipi\eta$ reaction study: 
The c.m. energy range ($\Ecm$), number of selected events after $\pipi\eta$-mass-resolution correction ($N_{\rm corr}$), 
detection efficiency ($\varepsilon$), 
differential ISR luminosity ($L$), 
and measured cross section ($\sigma$). 
The systematic uncertainty on the cross section in different energy intervals
is listed in Table~\ref{effcorr}.
\label{cross_table0}}
\begin{ruledtabular}
\begin{tabular}{ccccc |ccccc}
$\Ecm$, GeV & $N_{\rm corr}$ & $\varepsilon$, \% & L, nb$^{-1}$ & $\sigma$, nb 
& $\Ecm$, GeV & $N_{\rm corr}$ & $\varepsilon$, \% & L, nb$^{-1}$ & $\sigma$, nb\\
\hline
1.150 - 1.175 & $<$ 1 (90\% C.L.) & 1.36 & 1439.2 & $<$ 0.05 (90\% C.L.) & 1.875 - 1.900 & 86 $\pm$ 12 & 6.17 & 2430.7 & 0.575 $\pm$ 0.081 \\ 
1.175 - 1.20 & $<$ 1 (90\% C.L.) & 2.12 & 1468.3 &  $<$ 0.03 (90\% C.L.) & 1.900 - 1.925 & 136 $\pm$ 14 & 6.19 & 2468.7 & 0.888 $\pm$ 0.092 \\ 
1.20 - 1.225 & 9 $\pm$ 3 & 2.77 & 1498.0 & 0.231 $\pm$ 0.083 & 1.925 - 1.950 & 113 $\pm$ 13 & 6.18 & 2506.9 & 0.728 $\pm$ 0.086 \\ 
1.225 - 1.250 & 2 $\pm$ 2 & 3.33 & 1528.0 & 0.058 $\pm$ 0.052 & 1.950 - 1.975 & 115 $\pm$ 13 & 6.15 & 2545.2 & 0.736 $\pm$ 0.085 \\ 
1.250 - 1.275 & 13 $\pm$ 4 & 3.79 & 1558.6 & 0.228 $\pm$ 0.081 & 1.975 - 2.00 & 102 $\pm$ 12 & 6.08 & 2583.7 & 0.648 $\pm$ 0.081 \\ 
1.275 - 1.300 & 38 $\pm$ 7 & 4.18 & 1589.6 & 0.583 $\pm$ 0.112 & 2.00 - 2.05 & 138 $\pm$ 12 & 4.14 & 5283.5 & 0.632 $\pm$ 0.057 \\ 
1.300 - 1.325 & 32 $\pm$ 7 & 4.51 & 1621.1 & 0.444 $\pm$ 0.103 & 2.05 - 2.10 & 122 $\pm$ 11 & 4.14 & 5439.3 & 0.544 $\pm$ 0.050 \\ 
1.325 - 1.350 & 72 $\pm$ 10 & 4.77 & 1652.9 & 0.914 $\pm$ 0.134 & 2.10 - 2.15 & 78 $\pm$ 9 & 4.14 & 5596.4 & 0.337 $\pm$ 0.039 \\ 
1.350 - 1.375 & 107 $\pm$ 12 & 4.98 & 1685.1 & 1.280 $\pm$ 0.154 & 2.15 - 2.20 & 76 $\pm$ 9 & 4.14 & 5754.7 & 0.317 $\pm$ 0.038 \\ 
1.375 - 1.40 & 144 $\pm$ 15 & 5.15 & 1717.7 & 1.628 $\pm$ 0.170 & 2.20 - 2.25 & 58 $\pm$ 8 & 4.14 & 5914.1 & 0.236 $\pm$ 0.033 \\ 
1.400 - 1.425 & 195 $\pm$ 17 & 5.28 & 1750.7 & 2.103 $\pm$ 0.189 & 2.25 - 2.30 & 52 $\pm$ 7 & 4.14 & 6074.8 & 0.209 $\pm$ 0.031 \\ 
1.425 - 1.450 & 281 $\pm$ 20 & 5.38 & 1784.0 & 2.920 $\pm$ 0.216 & 2.30 - 2.35 & 82 $\pm$ 9 & 4.14 & 6236.7 & 0.317 $\pm$ 0.036 \\ 
1.450 - 1.475 & 357 $\pm$ 23 & 5.46 & 1817.6 & 3.582 $\pm$ 0.235 & 2.35 - 2.40 & 74 $\pm$ 9 & 4.14 & 6399.7 & 0.281 $\pm$ 0.033 \\ 
1.475 - 1.500 & 380 $\pm$ 24 & 5.53 & 1851.6 & 3.699 $\pm$ 0.237 & 2.40 - 2.45 & 60 $\pm$ 8 & 4.14 & 6564.1 & 0.223 $\pm$ 0.030 \\ 
1.500 - 1.525 & 419 $\pm$ 25 & 5.57 & 1885.9 & 3.970 $\pm$ 0.241 & 2.45 - 2.50 & 80 $\pm$ 9 & 4.14 & 6729.8 & 0.287 $\pm$ 0.032 \\ 
1.525 - 1.550 & 436 $\pm$ 26 & 5.61 & 1920.5 & 4.035 $\pm$ 0.240 & 2.50 - 2.55 & 49 $\pm$ 7 & 4.14 & 6897.0 & 0.173 $\pm$ 0.026 \\ 
1.550 - 1.575 & 424 $\pm$ 25 & 5.65 & 1955.3 & 3.826 $\pm$ 0.231 & 2.55 - 2.60 & 28 $\pm$ 5 & 4.14 & 7065.5 & 0.096 $\pm$ 0.019 \\ 
1.575 - 1.600 & 394 $\pm$ 24 & 5.68 & 1990.5 & 3.476 $\pm$ 0.218 & 2.60 - 2.65 & 44 $\pm$ 7 & 4.14 & 7235.7 & 0.147 $\pm$ 0.023 \\ 
1.600 - 1.625 & 355 $\pm$ 23 & 5.71 & 2025.9 & 3.065 $\pm$ 0.203 & 2.65 - 2.70 & 29 $\pm$ 5 & 4.14 & 7407.5 & 0.095 $\pm$ 0.018 \\ 
1.625 - 1.650 & 324 $\pm$ 22 & 5.74 & 2061.6 & 2.732 $\pm$ 0.189 & 2.70 - 2.75 & 30 $\pm$ 5 & 4.14 & 7581.0 & 0.097 $\pm$ 0.018 \\ 
1.650 - 1.675 & 307 $\pm$ 21 & 5.78 & 2097.5 & 2.528 $\pm$ 0.179 & 2.75 - 2.80 & 28 $\pm$ 5 & 4.14 & 7756.4 & 0.088 $\pm$ 0.017 \\ 
1.675 - 1.700 & 269 $\pm$ 20 & 5.82 & 2133.7 & 2.161 $\pm$ 0.166 & 2.80 - 2.85 & 33 $\pm$ 6 & 4.14 & 7933.8 & 0.101 $\pm$ 0.018 \\ 
1.700 - 1.725 & 285 $\pm$ 21 & 5.86 & 2170.1 & 2.233 $\pm$ 0.164 & 2.85 - 2.90 & 26 $\pm$ 5 & 4.14 & 8113.3 & 0.079 $\pm$ 0.015 \\ 
1.725 - 1.750 & 278 $\pm$ 20 & 5.91 & 2206.7 & 2.130 $\pm$ 0.159 & 2.90 - 2.95 & 15 $\pm$ 4 & 4.14 & 8294.9 & 0.044 $\pm$ 0.012 \\ 
1.750 - 1.775 & 280 $\pm$ 20 & 5.96 & 2243.6 & 2.091 $\pm$ 0.155 & 2.95 - 3.00 & 22 $\pm$ 5 & 4.14 & 8478.9 & 0.063 $\pm$ 0.014 \\ 
1.775 - 1.800 & 270 $\pm$ 20 & 6.01 & 2280.6 & 1.965 $\pm$ 0.149 & 3.00 - 3.05 & 20 $\pm$ 5 & 4.14 & 8665.4 & 0.058 $\pm$ 0.014 \\ 
1.800 - 1.825 & 282 $\pm$ 20 & 6.06 & 2317.9 & 2.005 $\pm$ 0.146 & 3.15 - 3.20 & 11 $\pm$ 4 & 4.14 & 9241.0 & 0.030 $\pm$ 0.010 \\ 
1.825 - 1.850 & 182 $\pm$ 17 & 6.11 & 2355.3 & 1.262 $\pm$ 0.118 & 3.20 - 3.30 & 26 $\pm$ 5 & 4.14 & 19077 & 0.033 $\pm$ 0.007 \\ 
1.850 - 1.875 & 145 $\pm$ 15 & 6.15 & 2392.9 & 0.987 $\pm$ 0.101 & 3.30 - 3.40 & 14 $\pm$ 4 & 4.14 & 19893 & 0.017 $\pm$ 0.005 \\ 
1.875 - 1.900 & 86 $\pm$ 12 & 6.17 & 2430.7 & 0.575 $\pm$ 0.081 & 3.40 - 3.50 & 7 $\pm$ 3 & 4.14 & 20737 & 0.008 $\pm$ 0.003 \\ 
\end{tabular}
\end{ruledtabular}
\end{table*}

\section{Detection efficiency and systematic uncertainties}
\begin{table}
\caption {Summary of the efficiency corrections and systematic uncertainties 
on the measured cross section.
\label{effcorr}}
\begin{ruledtabular}
\begin{tabular}{ccc}
Source        & Correction, \% & Systematic        \\
              &                & uncertainty, \%   \\
\hline
Selection criteria          &    & 2.5   \\
\hline
Background subtraction      &   &      \\
$m_{\pipi\eta}<1.35$        &   & 9  \\
$1.35<m_{\pipi\eta}<1.80$   &   & 2   \\
$1.80<m_{\pipi\eta}<2.50$   &   & 5  \\
$2.50<m_{\pipi\eta}<3.10$   &   & 10.5  \\
$3.10<m_{\pipi\eta}<3.50$   &   & 11   \\
\hline
Trigger and filters    & -1.5   & 1.6\\
$\eta$ reconstruction  & -2.0     & 1.0  \\
ISR photon efficiency  & -1.1   & 1.0 \\
Track reconstruction   & -1.1    & 1.0  \\
Radiative correction   &        & 1.0 \\
Luminosity             &        & 1.0 \\
\hline
Total                            &         &    \\
$m_{\pipi\eta}<1.35$        & -5.7  & 10  \\
$1.35<m_{\pipi\eta}<1.80$   & -5.7  & 4.5  \\
$1.80<m_{\pipi\eta}<2.50$   & -5.7  & 6.5  \\
$2.50<m_{\pipi\eta}<3.10$   & -5.7  & 11 \\
$3.10<m_{\pipi\eta}<3.50$   & -5.7  & 12   \\
\end{tabular}
\end{ruledtabular}
\end{table}
The corrected detection efficiency is defined as follows:
\begin{equation}\label{eff}
\varepsilon = \varepsilon_{\text{MC}} \prod_{i}(1+\delta_{i}),
\end{equation}
where $\varepsilon_{\text{MC}}$ is the detection efficiency
determined from MC simulation as the ratio of the true
$\pipi\eta$ mass spectrum obtained after applying
the selection criteria to the generated mass
spectrum, and $\delta_{i}$ are the efficiency corrections,
which take into account data-MC simulation differences in
track and photon reconstruction, $\chi^2_{\rm 4C}$ distribution,
etc. The detection efficiency $\varepsilon_{\text{MC}}$ as a function 
of  $E_{c.m.}$ is shown in Fig.~\ref{effmc} where the lines are fits to a fourth-order polynomial for $E_{c.m.} < 2$ GeV and 
to a constant for $E_{c.m.} > 2$ GeV. A discontinuity in the efficiency at 2 GeV is caused by additional
selection conditions used for Region II as mentioned in Sec.~\ref{event_select}.

To estimate efficiency corrections associated with the selection criteria, we
loosen a criterion, perform the procedure of background
subtraction described in the previous section, and calculate the ratio of
the number of selected events in data and simulation. 
For example, the condition $\chisq_{\rm 4C}<25(15)$ is loosened to
$\chisq_{\rm 4C}<300$. The efficiency correction is calculated 
as a relative difference between the data-MC simulation ratios calculated 
with the loosened and standard selection criteria. We do not observe any 
significant changes in data-MC simulation ratios due to variation of selection 
criteria and do not apply any corrections. 
The sum of the statistical uncertainties on the corrections for different 
selection criteria added in quadrature (2.5\%) is taken as an 
estimate of the systematic uncertainty associated with the selection criteria.

To estimate the uncertainty related to the description of the nonpeaking
background in the fit to the $m_{\gamma\gamma}$ spectrum, we repeat the fits using
a quadratic background. 
The main source of peaking background is the process $e^+e^-\to\pipi\piz\eta$.
Its contribution is calculated using the \jetsett $q\bar{q}$ simulation 
normalized as described in Sec.~\ref{event_select}. In the normalization we 
assume that \jetsett reproduces correctly the fraction of $\pipi\piz\eta$ events
in the full sample of $q\bar{q}$ events satisfying our selection criteria. To 
estimate the systematic uncertainty associated with this assumption, we vary the
fraction of $\pipi\piz\eta$ events by 50\%. The obtained uncertainties 
associated with the nonpeaking and peaking backgrounds added in quadrature are
listed in the section ``Background subtraction'' of Table~\ref{effcorr}.

We also study the quality of the simulation of the first-level trigger and 
background filters used in the primary event selection.
The overlap of the samples of events passing different filters and trigger 
selections is used to estimate the filter and trigger efficiency. The latter 
is found to be reproduced by simulation, with accuracy better than 
$5 \cdot 10^{-3}$. The correction due to data-MC simulation difference in the
filter inefficiency is determined to be $(-1.5 \pm 1.6)\%$.

To determine the efficiency correction for the data-MC simulation difference in
$\eta$ candidate reconstruction, we use the results of the study of the
$\piz$ reconstruction efficiency as a function of momentum described in Ref.~\cite{pi0gg_BaBar}. 
We assume that the $\eta\to\gamma\gamma$ efficiency is approximately
equal to the $\piz\to\gamma\gamma$ efficiency at the same energy, and
obtain the correction averaged over the $\eta$ momentum spectrum
$\delta_{\eta} = (-2 \pm 1) \%$. The correction is 
independent of the ${\pipi\eta}$ mass.

The ISR photon and charged-particle track reconstruction efficiencies are
studied in Ref.~\cite{4pic_BaBar}. The efficiency corrections and systematic uncertainties discussed in this section are summarized in Table~\ref{effcorr}. 

\begin{figure*}
\includegraphics[width=85mm]{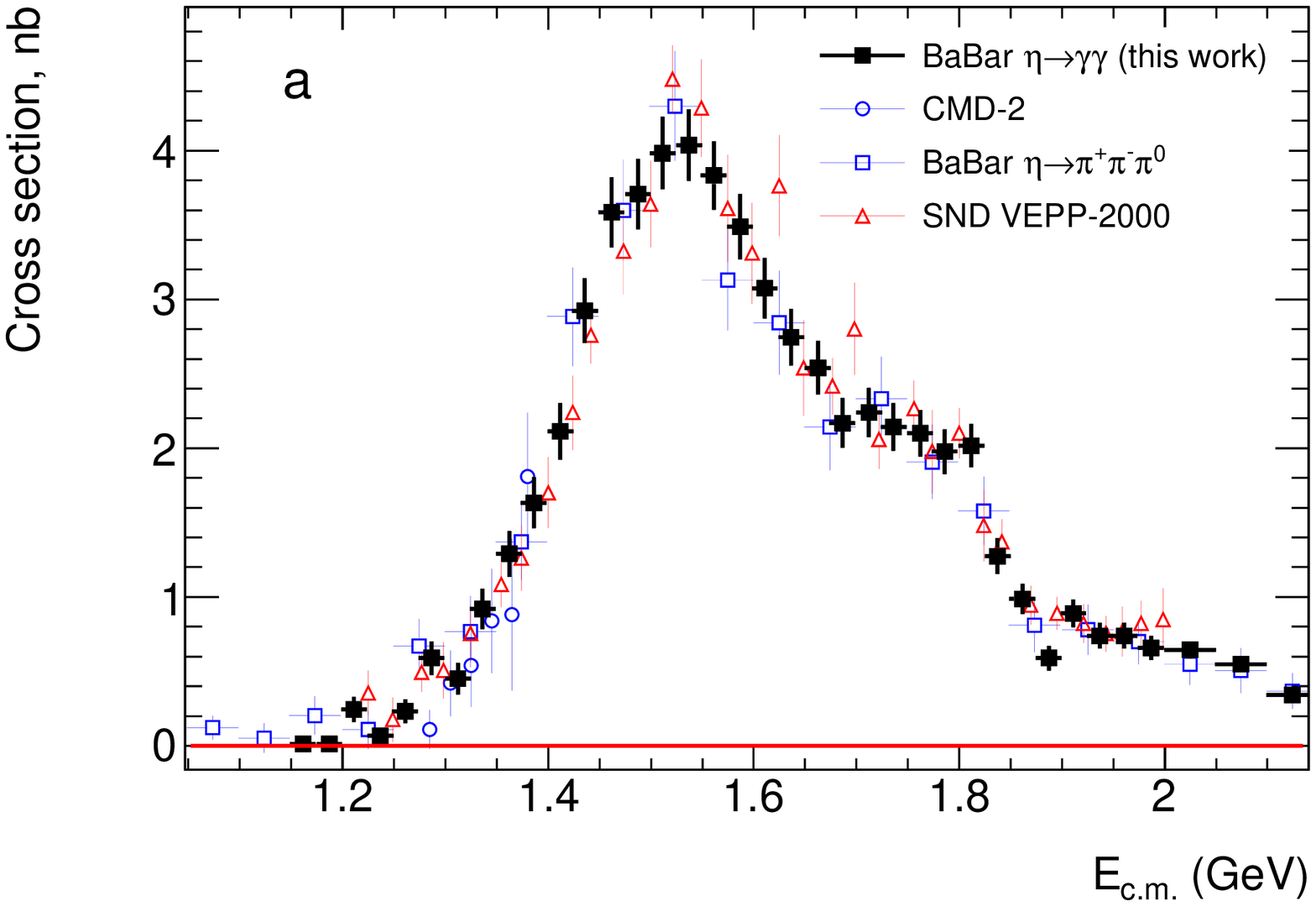}
\includegraphics[width=85mm]{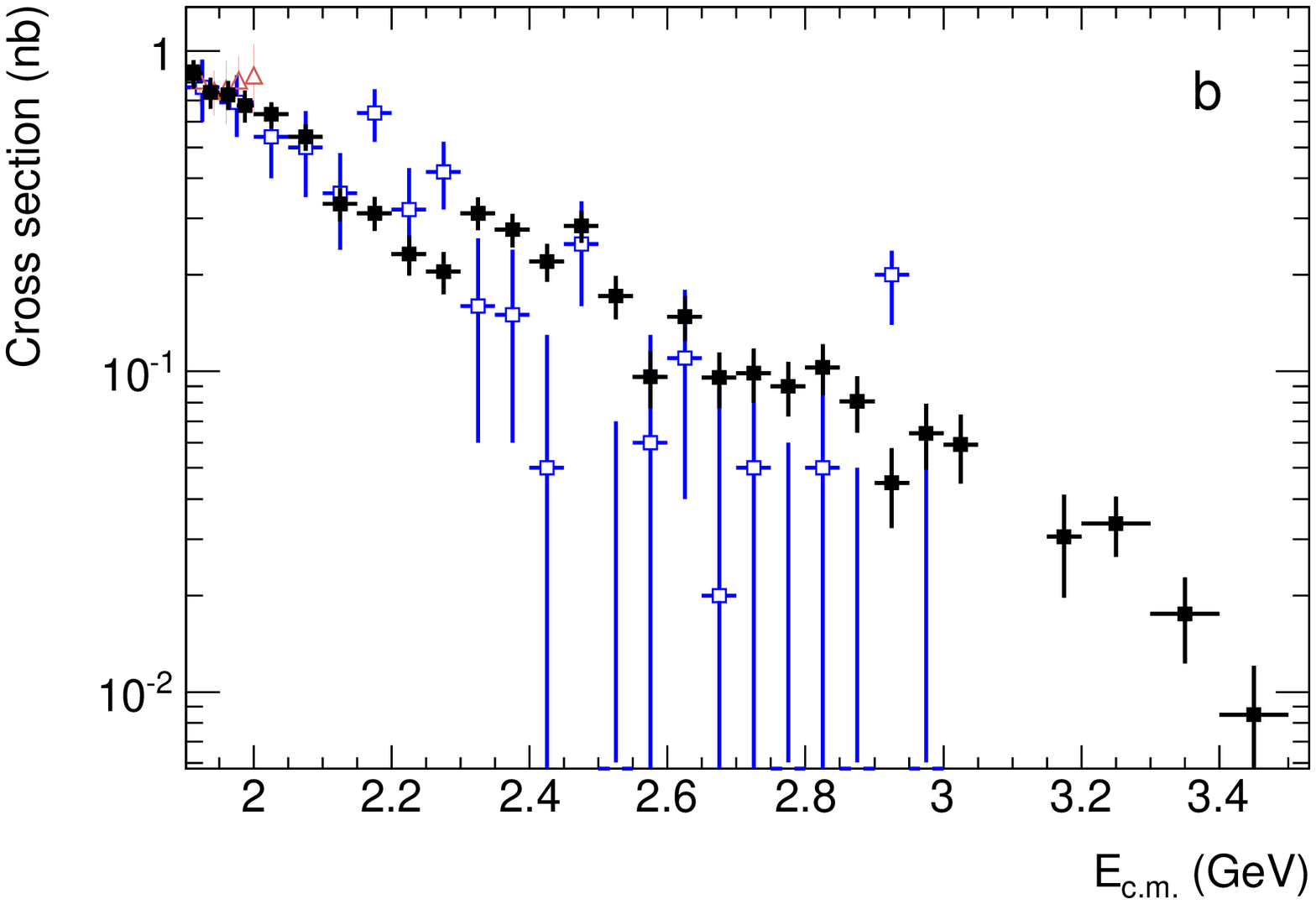}
\caption{ (color online) The $\epem \to \pipi\eta$ cross section measured in this work (\babar) 
in comparison with most precise previous measurements: CMD-2~\cite{CMD-2},
the independent \babar(2007)~\cite{BaBar2007}, SND~\cite{2pieta_SND2014}) at $\Ecm <2.15$ 
GeV (a) and $\Ecm >1.9$ GeV (b).}
\label{cross}
\end{figure*}

\section{The $\boldsymbol{ e^+ e^- \to \pi^+\pi^- \eta }$ cross section \label{xs2pieta}}
From the measured $\pipi\eta$ mass spectrum, we calculate the Born cross 
section
\begin{equation} \label{NumberEvents}
\sigma(m) = \frac{(dN/dm)_{\text{corr}}}{dL/dm\cdot\varepsilon(m) \cdot R} ~,
\end{equation}
where $m\equiv\Ecm/c^2$ is the invariant mass of the $\pipi\eta$ system, 
$(dN/dm)_{\text{corr}}$  is the $\pipi\eta$ mass spectrum after correction for the
detector mass resolution (unfolding),
$dL/dm$ is the so-called ISR differential luminosity~\cite{3piBaBar}, 
$\varepsilon(m)$ is the detection efficiency, and
$R$ is the radiative correction factor accounting for the Born $\pipi\eta$ mass
spectrum distortion due to emission of several photons
by the initial electron and positron. In our case the value of R is 
close to unity, and the theoretical uncertainty of $R$ does not 
exceed 1\%~\cite{NLO_ISR}. The uncertainty of the total integrated luminosity 
collected by \babar\ is less than 1\%~\cite{lumi}. 

The number of events in each bin $i$ ($N_i$) of the measured $\pipi\eta$ mass spectrum
shown in Fig.~\ref{Nevents} is related 
to the ``true'' number of events ($N_{\text{corr},i}$) as $N_i = \sum{A_{ij}N_{\text{corr},j}}$, 
where $A_{ij}$ is a migration matrix describing the probability for an
event with ``true'' mass in the bin $j$ to contribute to bin $i$.    
The matrix $A_{ij}$ is determined from the signal MC simulation.  
For the 25~\mev  bin width, diagonal elements of $A_{ij}$ are about 0.83,
and next-to-diagonal elements are about 0.08.
The inverse of the migration matrix is applied to the measured spectrum.
The obtained $(dN/dm)_{\text{corr}}$ spectrum is used to calculate the cross section. 
Since the cross section does not contain narrow structures, the unfolded mass
spectrum is close to the measured spectrum.
The differences between their bin contents are found to be less than half the statistical uncertainty. But the correction leads to an increase in the errors (by 4-15\%) and to correlations between the corrected numbers $N_{\text{corr},i}$. The neighbour to diagonal elements of the correlation matrix are about -20\% and the elements after next about 2\%.

The obtained $\epem\to\pipi \eta $ cross section 
is listed in Table~\ref{cross_table0} and shown in Fig.~\ref{cross} 
in comparison with the most precise previous measurements. The \babar(2007) results used a different
$\eta$ decay mode, and are independent.
The energy region near the $J/\psi$ resonance (3.05--3.15 GeV) is excluded 
from the data listed in Table~\ref{cross_table0} and is discussed below. The nonresonant cross 
section at $\Ecm = m_{J/\psi}$ will be obtained in Sec.~\ref{secpsi}.

Our cross section results are in agreement with previous measurements, have
comparable accuracy below 1.6 GeV and better accuracy above. 
In the energy range 3.0--3.5 GeV the cross sections are measured 
for the first time. 

\section{Fit to the $\boldsymbol{ e^+ e^- \to \pi^+\pi^- \eta }$ cross section\label{fit2pieta}}

In the framework of the VMD model the 
$\epem\to\pipi \eta $ cross section can be described by a coherent sum
of contributions from isovector states V that decay into $\rho(770)\eta$~\cite{NNAChasov}:
\begin{widetext}
\begin{equation}
\sigma(s) = \frac{4 \alpha^{2}}{3}\frac{1}{s\sqrt{s}} |F(s)|^{2} G(s),\,\,\, 
G(s) = \int_{4m_{\pi}^{2}}^{(\sqrt{s}-m_{\eta})^{2}}dq^{2} 
\frac{\sqrt{q^2}\Gamma_{\rho}(q^2)p_{\eta}^{3}(s,q^{2})}{(q^2-m_{\rho}^{2})^2
+(\sqrt{q^2}\Gamma_{\rho}(q^2))^2},
\label{modelcrosssection}
\end{equation}
\begin{equation}
p_{\eta}^2 = \frac{(s-m_{\eta}^2-q^2)^2 - 4m_{\eta}^2q^2}{4s},\,\,\,
\Gamma_{\rho}(q^2) = \Gamma_{\rho}(m_{\rho}^2)\frac{m_{\rho}^2}{q^2}\left(\frac{p^2_{\pi}(q^2)}{p^2_{\pi}(m_{\rho}^2)}\right)^{\frac{3}{2}},\,\,\,
p^2_{\pi}(q^2) = q^2/4 - m_{\pi}^2,
\end{equation}
\end{widetext}
where $\sqrt{s} = E_{c.m.}$, $q$ is the $\pipi$ invariant mass,
$m_{\eta}$ and $m_{\pi}$ are the $\eta$ meson and charged pion masses,
$m_{\rho}$ and $\Gamma_{\rho}(m_{\rho}^2)$ are the $\rho(770)$ mass
and width, and
\begin{equation}
F(s) = \sum_{V} \frac{m_{V}^2g_Ve^{i\phi_{V}}}
{s-m_{V}^2+i\sqrt{s}\Gamma_{V}(s)},
\label{modelformfactor}
\end{equation}
where the sum is over all $\rho$ resonances and
the complex parameter $g_{V}e^{i\phi_{V}}$ is the 
combination $g_{V\rho\eta}/g_{V\gamma}$
of the coupling constants describing the transitions 
$V \to \rho \eta$ and $V \to \gamma^{\star}$, respectively.

The VMD model [Eq.(\ref{modelcrosssection})] is used to fit our cross section
data. The free fit parameters are $g_{V}$, and the masses and widths of the 
excited $\rho$-like states. The $\rho(770)$ mass and width are fixed at their 
Particle Data Group (PDG) values~\cite{PDG}. The phase $\phi_{\rho(770)}$ is 
set to zero. The coupling constants $g_{V\rho\eta}$ and $g_{V\gamma}$ are not 
expected to have sizable imaginary parts~\cite{2pieta_SND2014}. Therefore, 
we assume that $\phi_{V}$ for the excited states are 0 or $\pi$.

\begin{figure*}
\begin{center}
\includegraphics[width=120mm]{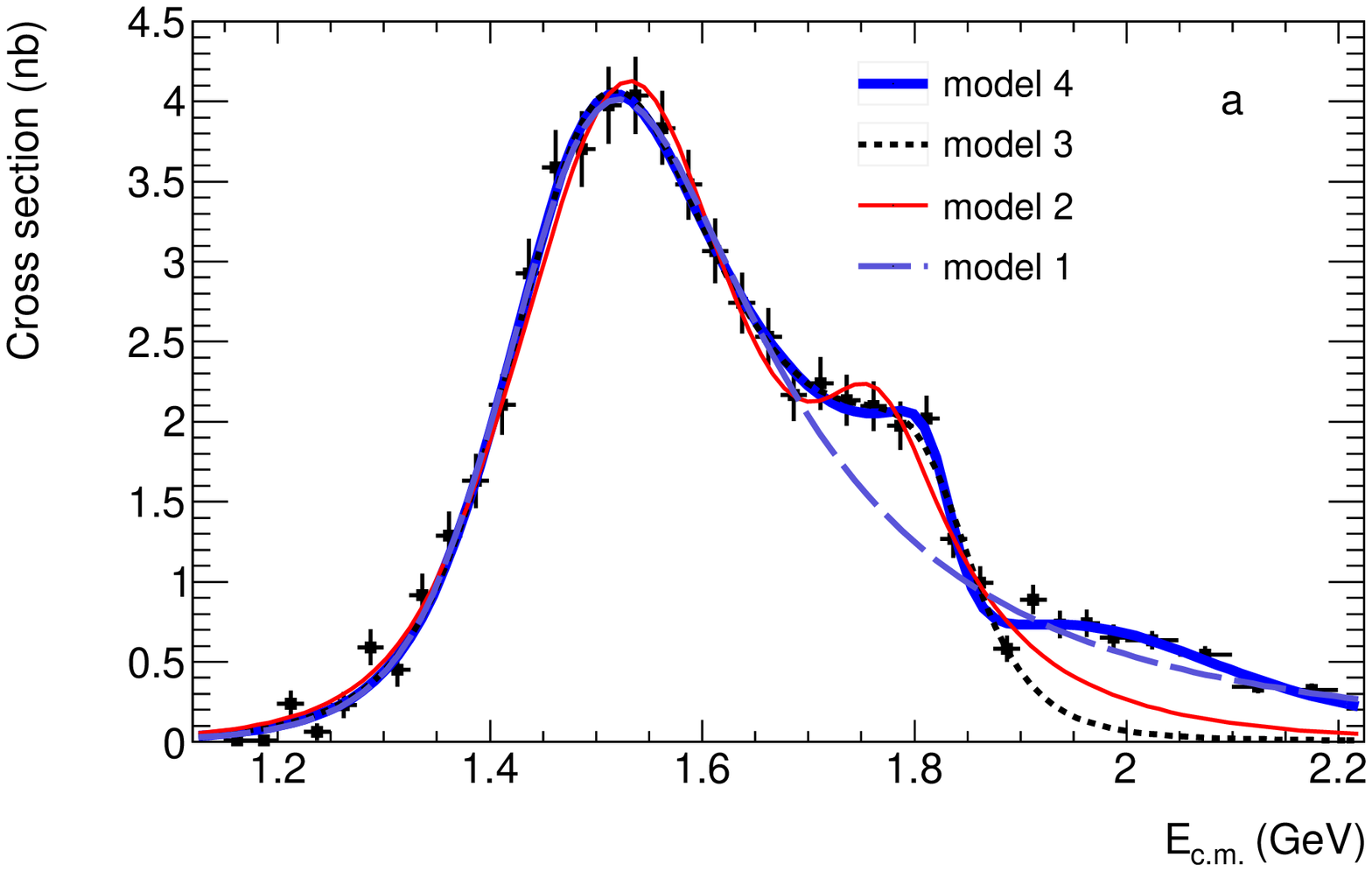} 
\includegraphics[width=120mm]{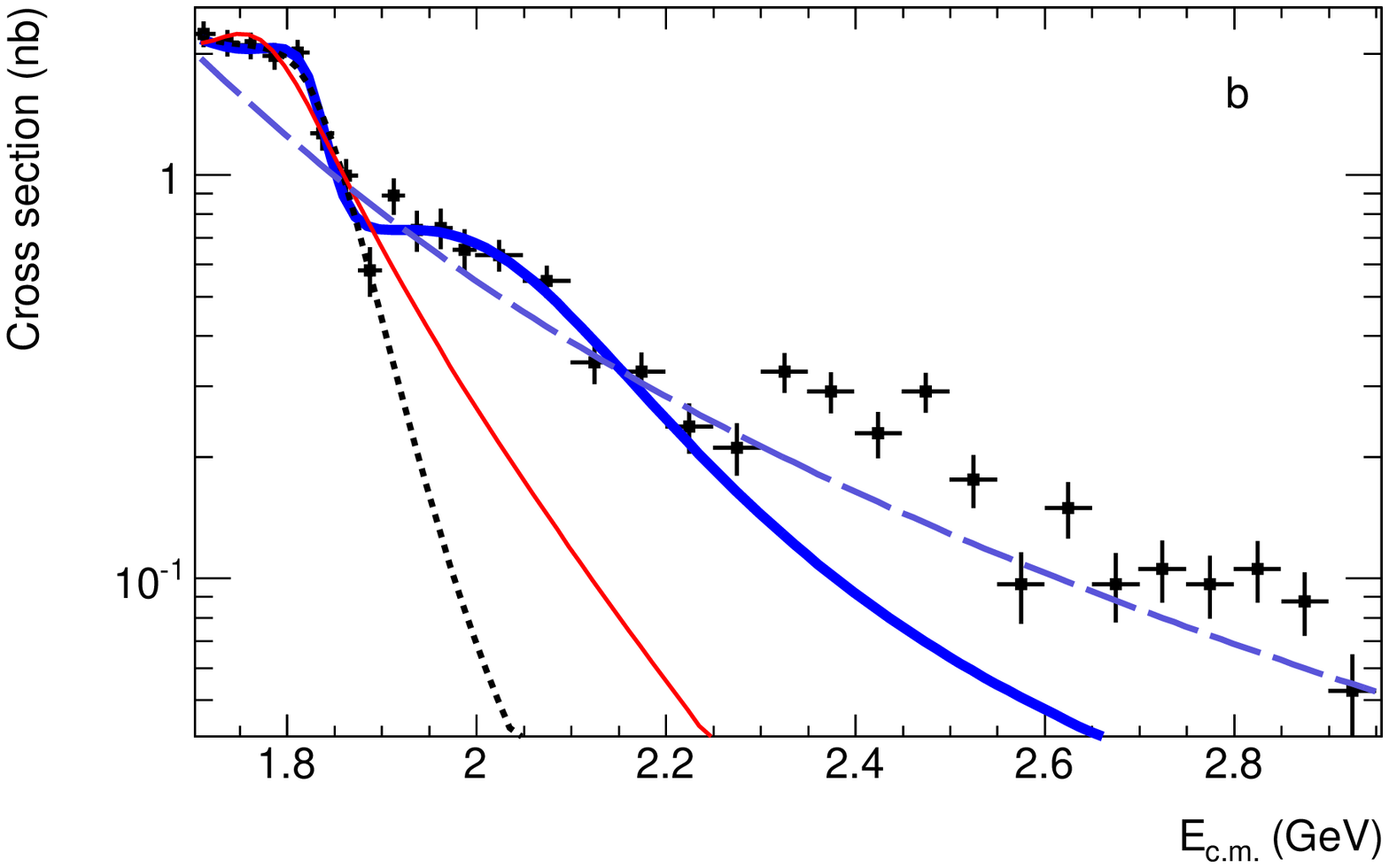} 
\caption{(color online) The measured $\epem\to\pipi \eta $ cross section fitted
with the four models described in the text.\label{cross_fit}}
\end{center}
\end{figure*}

The models with one, two, and three excited states are tested.
In Model 1, the cross section data are fitted in the energy range
$\Ecm = 1.2-1.70$~\gev with two resonances,
$\rho(770)$ and $\rho(1450)$.
The model with $\phi_{\rho(1450)}=0$ fails to describe the data.
The fit result with $\phi_{\rho(1450)}=\pi$ is shown in Fig.~\ref{cross_fit}
by the long-dashed curve. The obtained fit parameters are listed in 
Table~\ref{crossapprox_1}. 
It is seen that Model 1 cannot reproduce
the structure in the cross section near 1.8 GeV. 
\begin{table*}
\caption { The coupling constants and resonance parameters obtained in 
the fits to the $\epem\to\pipi\eta$ cross section data. \label{crossapprox_1}}
\begin{ruledtabular}
\begin{tabular}{ccccc}
    Parameter                  & Model 1          & Model 2           & Model 3 & Model 4 \\%
\hline    
$g_{\rho(770)}$, GeV$^{-1}$    & $1.1\pm 0.3$     & $2.3\pm 0.3$      & $1.8 \pm  0.3$   & $1.7\pm 0.3$    \\
$g_{\rho(1450)}$, GeV$^{-1}$   & $0.49\pm 0.02$   & $0.36\pm 0.05$    & $0.44\pm  0.02$  & $0.46\pm 0.03$  \\
$g_{\rho(1700)}$, GeV$^{-1}$   & --               & $0.044\pm 0.019$  & $0.080\pm 0.012$ & $0.016\pm 0.007$\\
$g_{\rho'''}    $, GeV$^{-1}$  & --               & --                & --               & $0.09\pm 0.02$  \\
$m_{\rho(1450)}$, \gevcc       &$1.487\pm 0.016$  & $1.54\pm 0.01$    & $1.50 \pm 0.01$  & $1.49\pm 0.01$  \\
$m_{\rho(1700)}$, \gevcc       & --               & $1.76\pm 0.01$    & $1.83 \pm 0.01$  & $1.83\pm 0.01$  \\
$m_{\rho'''}   $, \gevcc       & --               & --                & --          	 & $2.01\pm 0.04$  \\												  
$\Gamma_{\rho(1450)}$, GeV     &$0.33\pm 0.02$    & $0.31\pm 0.03$    & $0.28 \pm 0.02$  & $0.29\pm 0.02$  \\
$\Gamma_{\rho(1700)}$, GeV     & --               & $0.16\pm 0.04$    & $0.17 \pm 0.02$	 & $0.08\pm 0.02$  \\
$\Gamma_{\rho'''}   $, GeV     & --               & --                & --       	     & $0.42\pm 0.09$  \\
$\phi_{770,1450;...}$         & 0; $\pi$            & 0; $\pi; ~\pi$         & 0; $\pi$; 0    & 0; $\pi$; 0; 0       \\
 $\chi^2$ per d.o.f.           & 14/16            & 35/21             & 19/21            & 28/26           \\
\end{tabular}
\end{ruledtabular}
\end{table*}

In Models 2 and 3 we include an additional contribution from the $\rho(1700)$ 
resonance with phases $\phi_{\rho(1700)}$ = $\pi$ and 0, respectively.
The fits are done in the range $\Ecm =$ 1.2--1.90~\gev. 
The fit results are shown in Fig.~\ref{cross_fit} and listed
in Table~\ref{crossapprox_1}. Both models describe the data below 1.90 \gev
reasonably well. Model 3 has better $\chi^2$ ($P(\chi^2)=0.58$ instead of $0.03$ for Model 2). 
Above 1.90 \gev the fit curves for both the models 
lie below the data.

Model 4 is Model 3 with a fourth resonance $\rho'''$ added. 
The phase $\phi_{\rho'''}$ is set to zero. 
The fitted 
energy range is extended up to 2.2 \gev. The fit result is shown 
in Fig.~\ref{cross_fit}. The fitted resonance mass $m_{\rho'''}=2.01\pm0.04$ GeV is between
the masses of the $\rho(1900)$ and $\rho(2150)$ states
listed in the PDG table~\cite{PDG}. 
The fitted value $g_{\rho} = 1.7 \pm0.3$ GeV$^{-1}$ agrees
with the VMD estimation of $1.57 \pm 0.07$ GeV$^{-1}$ from the partial width $\rho(770) \to \eta\gamma$.
It is seen that the model successfully describes 
the cross section data up to  2.3 \gev. Above $\Ecm =$ 2.3 \gev Model 4 lies below the data,
which could be explained by another resonance. Alternatively, the change of the cross section slope near
1.9 GeV may be interpreted without inclusion of a fourth resonance,
as a threshold effect due to the opening of the nucleon-antinucleon
production channel. Structures near the nucleon-antinucleon threshold 
are observed in the $e^+e^- \to 3(\pi^+\pi^-)$ and
$2(\pi^+\pi^-\pi^0)$ cross sections~\cite{threshold1,threshold2} as well as in the $\eta'\pi^+\pi^-$ mass spectrum in the decay $J/\psi \to \gamma \eta'\pi^+\pi^-$~\cite{threshold4}. 
A slope change near 1.9 GeV is seen in the $e^+e^- \to \pi^+\pi^-\pi^+\pi^-$
cross section~\cite{threshold3}.

The fit is also performed with another parametrization. The parameters
$g_{V}$ are replaced by the products 
\begin{equation} 
\Gamma(V\to e^+e^-)\BR(V\to\eta\pi^+\pi^-) =
\frac{\alpha^{2}}{9\pi} \frac{|g_{V}|^{2}m_{V}}{\Gamma_{V}}
G(m_{V}^{2}).
\end{equation}
From the fit in Model 3 we obtain:
\begin{eqnarray}
\Gamma(\rho(1450)\to\epem)\BR(\rho(1450)\to\eta\pipi)=\nonumber\\
(210 \pm 24_{\text{stat}} \pm 10_{\text{syst}}) \rm~ eV\nonumber\\
\Gamma(\rho(1700)\to\epem)\BR(\rho(1700)\to\eta\pipi)=\nonumber\\
 (84 \pm 26_{\text{stat}} \pm 4_{\text{syst}}) \rm~ eV
\end{eqnarray}
The model uncertainties of these
parameters estimated from the difference of fit results for Model 2, 3, and 4,
are large, 20\% for $\rho(1450)$ and 80\% for $\rho(1700)$.

\section{Test of CVC}
The CVC hypothesis and isospin symmetry
allow the prediction of the $\pi^{-}\pi^0 \eta$ mass spectrum and the 
branching fraction for the $\tau^-\to\pi^-\piz\eta \nu_{\tau}$ decay 
from data for the $\epem \to \pipi \eta $ cross section~\cite{CVC}.
The branching fraction can be calculated as:
\begin{equation}
\begin{gathered}
\frac{\BR(\tau^{-} \to \pi^{-}\piz \eta \nu_{\tau})}{\BR(\tau^{-} \to e^{-} \bar{\nu_e} 
\nu_{\tau})}
= \int_{(2m_{\pi}+m_{\eta})^2}^{m^2_{\tau}} dq^2  \\
\sigma^{I=1}_{e^+e^- \to \pi^+ \pi^- \eta}(q^2)
\frac{3|V_{ud}|^{2}S_{\text{EW}}}{2\pi\alpha^2} 
\frac{q^2}{m_{\tau}^2}
(1-\frac{q^2}{m_{\tau}^2})^2
(1+2\frac{q^2}{m_{\tau}^2}), 
 \label{BBB}
\end{gathered}
\end{equation} 
where $q^{2}$ is the squared 4-momentum of the $\pi^{\pm}\piz \eta$ system, 
$|V_{ud}|$ is the Cabibbo-Kobayashi-Maskawa matrix element, and
$S_{\text{EW}} = 1.0194$ is a factor taking into account electroweak radiative 
corrections, and $\BR(\tau^{-} \to e^{-} \bar{\nu_e} \nu_{\tau})$ = 17.83 $\pm$ 0.04\%~\cite{PDG}. 

We integrate Eq.(\ref{BBB}) using the fit function for the cross
section  of model $\#4$ from the previous section and obtain
\begin{equation}
\begin{gathered}
\BR(\tau^{-} \to \pi^{-}\pi^0 \eta \nu_{\tau}) =  (0.1616 \pm 0.0026_{\text{stat}}\pm \\
0.0080_{\text{syst}} \pm 0.0011_{\text{model}}) \% = (0.162 \pm 0.009) \%,
\label{iniiiiininn}
\end{gathered}
\end{equation} 
where the first error is statistical, the second is systematic 
(see Table~\ref{effcorr}), and the third is model uncertainty.

The latter is estimated from the 
difference between the branching fraction values obtained
with the cross section parametrization in Model 2 and Model 3
discussed in the previous section.
\begin{figure*}
\includegraphics[width=85mm]{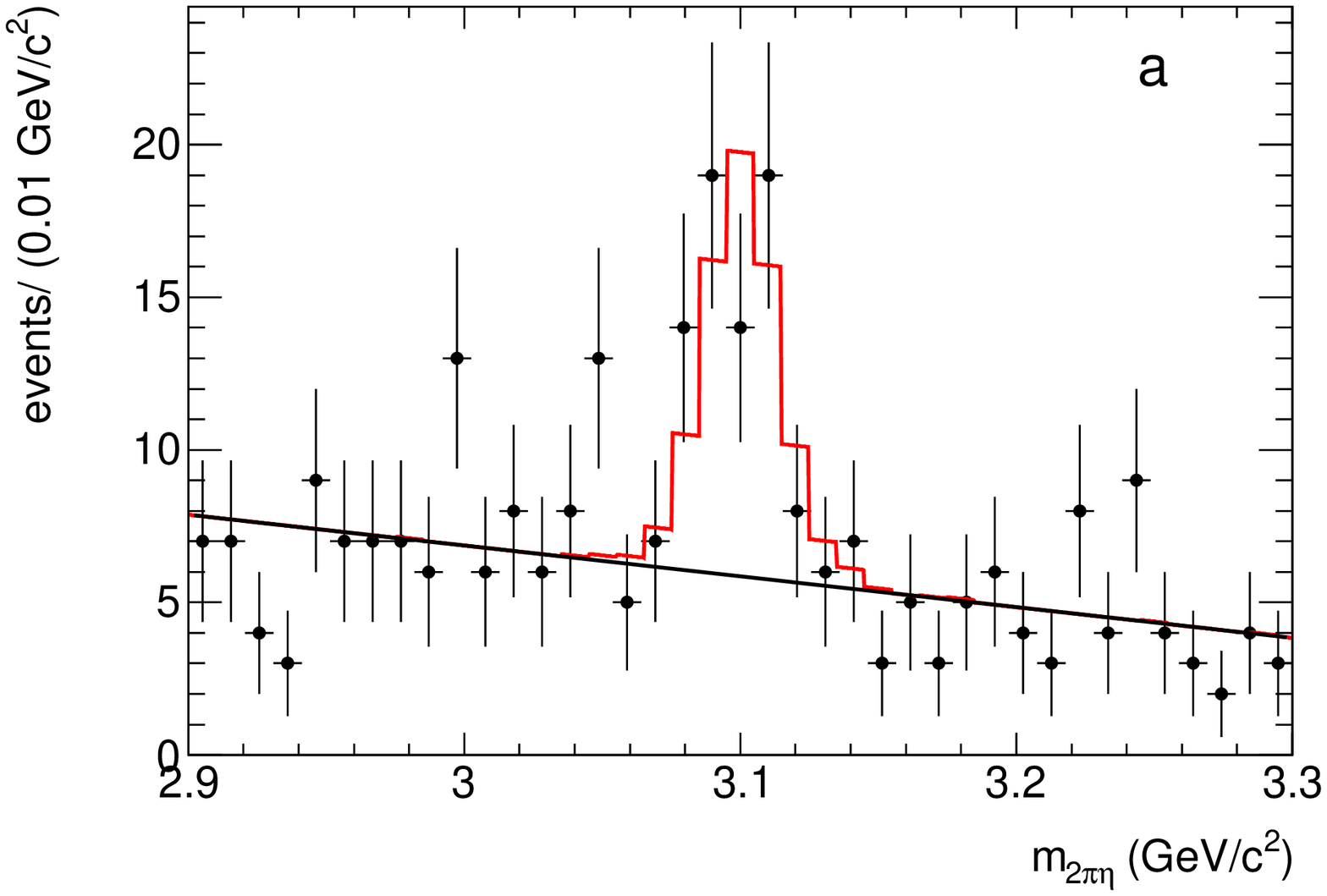}
\includegraphics[width=85mm]{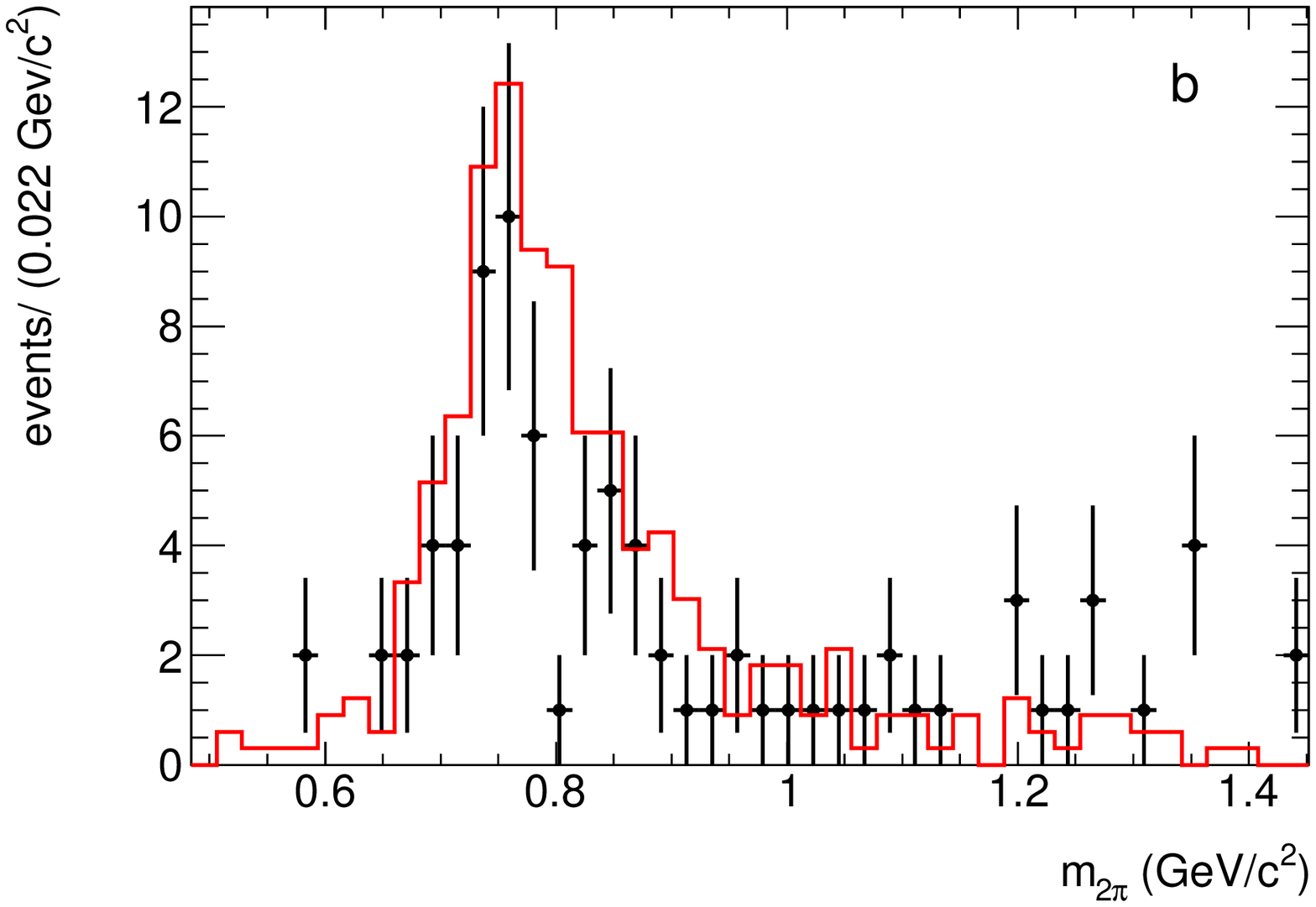}
\caption{(a) The fit to the $\pipi\eta$ mass spectrum for data events from
the $J/\psi$ region.
(b) The $m_{\pipi}$ invariant mass distribution for
data events with $3.05 < m_{\pipi\eta} < 3.15 $ GeV/c$^2$ (points with error 
bars) and simulated signal events generated using the model with the 
$\rho(770)\eta$ intermediate state (histogram).
\label{J_psi}}
\end{figure*}
The calculation based on the previous \babar\ measurement of the 
$\pipi\eta\to\pipi\pipi\piz$ final state~\cite{BaBar2007} gives
$\BR(\tau^{\pm} \to \pi^{\pm}\piz \eta \nu_{\tau}) = (0.1695 \pm 0.0085_{\text{stat}} 
\pm 0.0136_{\text{syst}}) \%$, compatible with the new result~(\ref{iniiiiininn}). The systematic uncertanties on the luminosity, 
radiative corrections, photon and track efficiencies are the same for the
new and previous \babar\ measurements.  
Combining the two \babar\ values we obtain
\begin{equation}
\BR(\tau^{-} \to \pi^{-}\piz \eta \nu_{\tau}) = (0.163 \pm 0.008)\%, 
\end{equation}
which is in good agreement with, but more precise than, the estimate based 
on the SND $\epem\to\pipi \eta $ measurement
$(0.156 \pm 0.011)\%$~\cite{2pieta_SND2014}.

The PDG value of this branching fraction is
$\BR(\tau^- \to \pi^-\pi^0 \eta \nu_{\tau})_{\rm exp} =
(0.139 \pm 0.010)\%$~\cite{PDG}. The difference between the experimental result
and our CVC-based calculation is $1.8\sigma$. The difference, 
about 15\% of the branching fraction, is too large to be explained by 
isospin-breaking corrections. The quoted PDG value is based on the three 
measurements:
$(0.135 \pm 0.003 \pm 0.007)\%$ by Belle~\cite{Belle2009},
$(0.18 \pm 0.04 \pm 0.02)\%$ by ALEPH~\cite{ALEPH1997}, and
$(0.17 \pm 0.02 \pm 0.02)\%$ by CLEO~\cite{CLEO1992}. Its error includes
a scale factor of 1.4.
The difference between our CVC prediction and the most precise measurement
by Belle is $2.4\sigma$.

\section{The $\boldsymbol{ J/\psi \to \pi^+\pi^- \eta }$ decay\label{secpsi}}
The $\pipi\eta$ mass spectrum for selected data events in the region 
near the $J/\psi$ is shown in Fig.~\ref{J_psi}(a). The spectrum is fitted
by a sum of a function describing the $J/\psi$ line shape and 
a linear background function. The $J/\psi$ line shape is obtained using MC 
simulation. The fit yields $49 \pm 9$ events of the decay $J/\psi\to\pipi\eta$.

From the fitted number of $J/\psi$ events we calculate the 
product~\cite{NLO_ISR}
\begin{equation}
\begin{gathered} \label{Jpsi}
\Gamma(J/\psi \to e^{+}e^{-}) \BR(J/\psi \to \pipi\eta) =\\ 
\frac{N_{J/\psi}m_{J/\psi}^2}{6\pi^2dL/dm(m_{J/\psi})\varepsilon(m_{J/\psi})} =\\
 (2.34 \pm 0.43_{\text{stat}} \pm 0.16_{\text{syst}})~\rm eV. 
\end{gathered}
\end{equation} 
Using the nominal value of the $J/\psi$ electron width 
$(5.55 \pm 0.14)$ eV~\cite{PDG} we obtain the branching fraction
\begin{equation}\label{sdfsdfsdf}
\BR(J/\psi\to \pipi\eta) = (4.2 \pm 0.8)\times10^{-4},
\end{equation}
which has better 
precision than the current PDG value $(4.0 \pm 1.7)\times10^{-4}$~\cite{PDG}.

Figure~\ref{J_psi}(b) shows the $m_{\pipi}$ invariant mass distributions for
data events from the $J/\psi$ peak ($3.05 < m_{\pipi\eta} < 3.15 $ GeV/c$^2$)
and simulated events. The simulation uses the model with the $\rho(770)\eta$ 
intermediate state. The difference between the $m_{\pipi}$ distributions for data and simulation
is explained by the contribution of the isoscalar $\omega\eta$ intermediate
state and its interference with the isovector 
amplitudes, where $\rho(770)\eta$ gives the main contribution~\cite{J_psi_VP_DM2, J_psi_VP_MARK3}.

The G-parity of $\pipi \eta$ is +1, whereas G($J/\psi$) = -1.
Therefore, this final state cannot be reached in strong-interaction
(``direct'') decays. An allowed way for the decay is electromagnetic,
$J/\psi \to \gamma^* \to \pi^+ \pi^- \eta$. If this is the only way, the
branching fraction has to fulfill:
\begin{equation}\label{jpsii}
\begin{gathered} 
 \BR(J/\psi \to \pi^+ \pi^- \eta) / \BR(J/\psi \to \mu^+ \mu^-) = \\
     \sigma_c (e^+ e^- \to \pi^+ \pi^- \eta) / \sigma_c (e^+ e^- \to \mu^+\mu^-),
\end{gathered}
\end{equation}
where $\sigma_c$ is the continuum cross section at $\sqrt{s} = m_{J/\psi}$ and
$\sigma_c (e^+ e^- \to \mu^+ \mu^-) = \frac{4 \pi \alpha^2}{3 m^2_{J/\psi}}$.

We obtain the continuum cross section for $\pi^+ \pi^- \eta$ production by
linear interpolation between four points near $m_{J/\psi}$, where two lie
below 3.05 GeV/c$^2$ and two above 3.15 GeV/c$^2$:
\begin{equation}
\sigma_c (e^+ e^- \to \pi^+ \pi^-  \eta)=(47\pm 8_{\text{stat}}\pm 5_{\text{syst}}) \mbox{pb}.
\end{equation}

Inserting this result into Eq.(\ref{jpsii}) leads to
\begin{equation}
\begin{gathered} 
 \BR (J/\psi \to \pi^+ \pi^- \eta) = \frac{3 m^2_{J/\psi}}{4 \pi \alpha^2} 
           \BR(J/\psi \to \mu^+ \mu^-) \\ \times \sigma_c (e^+ e^- \to \pi^+ \pi^- \eta)
                          = (3.1 \pm 0.6) \times 10^{-4}.
\end{gathered}
\end{equation}

This is smaller than the result in Eq.(\ref{sdfsdfsdf}) by $(1.1 \pm 1.0) \times 10^{-4}$.
A second way to violate G-parity is the direct decay $J/\psi \to \omega \eta$
followed by the G-violating decay $\omega \to \pi^+ \pi^-$. Our result
confirms that there could be a sizeable contribution of the
$\omega \eta$ intermediate state to the decay $J/\psi \to \pi^+ \pi^- \eta$. 

\section{Summary}
In this paper we have studied the process $\epem \to \pipi\eta\gamma$, 
in which the photon is emitted from the initial state.
Using the ISR technique we have measured the $\epem\to\pipi\eta$ cross section 
in the c.m. energy range from 1.15 up to 3.5 GeV. 
Our results are in agreement with previous measurements, including our
 own previous result in the independent $\eta\to\pip\pim\piz$ channel, and have
comparable precision below 1.6 GeV and better precision above.
In the energy range below 2.2 GeV the measured cross section is well described
by the VMD model with four $\rho$-like resonances. Parameters of these
resonances have been obtained.

Using the measured cross section and the CVC hypothesis, the branching 
fraction of the decay $\tau^{-} \to \eta \pi^{-} \pi^{0} \nu_{\tau}$ is 
determined to be 
$\BR(\tau^{-} \to \pi^{-}\pi^0 \eta \nu_{\tau}) = (0.162 \pm 0.009)$ \%. 

From the measured number of $\epem \to J/\psi\gamma \to \pipi \eta \gamma$
events we have determined the product 
$\Gamma_{J/\Psi \to e^{-}e^{+}} \BR_{J/\psi \to \pipi\eta} = 2.34 \pm 0.46 $ eV,
and the branching fraction $\BR(J/\psi\to \pipi\eta) = (0.042 \pm 0.008) \%$.

\section{ACKNOWLEDGMENTS}

We are grateful for the 
extraordinary contributions of our \pep2\ colleagues in
achieving the excellent luminosity and machine conditions
that have made this work possible.
The success of this project also relies critically on the 
expertise and dedication of the computing organizations that 
support \babar.
The collaborating institutions wish to thank 
SLAC for its support and the kind hospitality extended to them. 
This work is supported by the
US Department of Energy
and National Science Foundation, the
Natural Sciences and Engineering Research Council (Canada),
the Commissariat \`a l'Energie Atomique and
Institut National de Physique Nucl\'eaire et de Physique des Particules
(France), the
Bundesministerium f\"ur Bildung und Forschung and
Deutsche Forschungsgemeinschaft
(Germany), the
Istituto Nazionale di Fisica Nucleare (Italy),
the Foundation for Fundamental Research on Matter (The Netherlands),
the Research Council of Norway, the
Ministry of Education and Science of the Russian Federation, 
Ministerio de Econom\'{\i}a y Competitividad (Spain), the
Science and Technology Facilities Council (United Kingdom),
and the Binational Science Foundation (U.S.-Israel).
Individuals have received support from 
the Marie-Curie IEF program (European Union) and the A. P. Sloan Foundation (USA). 



\newpage  
\begin{thebibliography}{0}

\bibitem{NLO_ISR} M. Benayoun {\it et al.}, 
Mod. Phys. Lett. A {\bf 14}, 2605 (1999). 

\bibitem{ISRprinciple} A. B. Arbuzov {\it et al.},
JHEP {\bf 9812}, 009 (1998).

\bibitem{ISRprinciple1} S. Binner, J. H. K{\"u}hn, and K. Melnikov, 
Phys. Lett. B {\bf 459}, 279 (1999).

\bibitem{NNAChasov}N. N. Achasov and V. A. Karnakov, 
JETP Lett. {\bf39}, 285 (1984).

\bibitem{CVCEidelman}V. A. Cherepanov and S. I. Eidelman, 
JETP Lett. {\bf89}, 9 (2009).

\bibitem{DM1} A. Cordier {\it et al.} (DM1 Collaboration), 
Nucl. Phys. B {\bf172}, 13 (1980).

\bibitem{ND} V. P. Druzhinin {\it et al.} (ND Collaboration), 
Phys. Lett. B {\bf174}, 115 (1986).

\bibitem{DM2} A. Antonelli {\it et al.} (DM2 Collaboration), 
Phys. Lett. B {\bf212}, 133 (1988).

\bibitem{CMD-2} R. R. Akhmetshin {\it et al.} (CMD-2 Collaboration), 
Phys. Lett. B {\bf489}, 125 (2000).

\bibitem{2pieta_SND} M. N. Achasov {\it et al.} (SND Collaboration),
JETP Lett. {\bf92}, 80 (2010).

\bibitem{2pieta_SND2014} V. M. Aulchenko {\it et al.} (SND Collaboration), 
Phys. Rev. D {\bf91}, 052013 (2015).

\bibitem{BaBar2007}
B. Aubert {\it et al.} (\babar ~Collaboration), 
Phys. Rev. D. {\bf76}, 092005 (2007).

\bibitem{NJL}M. K. Volkov {\it et al.}, 
Phys. Rev. C {\bf89}, 015202 (2014).

\bibitem{ResonChirTh}D. G. Dumm {\it et al.},
Phys. Rev. D {\bf86}, 076009 (2012).

\bibitem{lumi}J. P. Lees {\it et al.} (\babar ~Collaboration), 
Nucl. Instrum. Methods Phys. Res., Sect. A {\bf726}, 203 (2013).

\bibitem{Detector}B. Aubert {\it et al.} (\babar ~Collaboration), 
Nucl. Instrum. Methods Phys. Res., Sect. A {\bf 479}, 1 (2002).

\bibitem{Detector1}B. Aubert {\it et al.} (\babar ~Collaboration),
Nucl. Instrum. and Meth. A {\bf 729}, 615 (2013).

\bibitem{EVA}H. Czy$\dot{\rm z}$ and J. H. K$\ddot{\rm u}$hn, 
Eur. Phys. J. C {\bf 18}, 497 (2001).

\bibitem{structuremethod}M. Caffo, H. Czy$\dot{\rm z}$, E. Remiddi, 
Nuo. Cim. A {\bf{110}}, 515 (1997); Phys. Lett. B {\bf 327}, 369 (1994).

\bibitem{photos}E. Barberio, B. van Eijk and Z. Was, 
Comput. Phys. Commun. {\bf 66}, 115 (1991).

\bibitem{GEANT4} S. Agostinelli {\it et al.} (Geant4 Collaboration), 
Nucl. Instrum. Methods Phys. Res., Sect. A {\bf 506}, 250 (2003).

\bibitem{udssim}
T. Sj\"{o}strand, Comput. Phys. Commun. {\bf 82}, 74 (1994).

\bibitem{2pi2pi0datababar}
J. P. Lees {\it et al.} (\babar ~Collaboration), Phys. Rev. D {\bf 96}, 092009  (2017). 

\bibitem{babar2keta}
B. Aubert {\it et al.} (\babar ~Collaboration), 
Phys. Rev. D {\bf 77}, 092002 (2008).

\bibitem{snd3pieta}
R. R. Akhmetshin {\it et al.},  Phys. Lett. B {\bf773}, 150 (2017).

\bibitem{tautausim} 
S. Jadach and Z. Was, Comput. Phys. Commun. {\bf 85}, 453 (1995).

\bibitem{3piBaBar}
B. Aubert {\it et al.} (\babar ~Collaboration), 
Phys. Rev. D {\bf70}, 072004 (2004).

\bibitem{pi0gg_BaBar} B. Aubert {\it et al.} (\babar ~Collaboration), 
Phys. Rev. D {\bf80}, 052002 (2009).

\bibitem{4pic_BaBar} J. P. Lees {\it et al.} (\babar ~Collaboration),
Phys. Rev. D {\bf85}, 112009 (2012).

\bibitem{CVC} Y. S. Tsai, Phys. Rev. D {\bf4}, 2821 (1971).

\bibitem{threshold1}  R. R. Akhmetshin {\it et al.} (CMD-3 Collaboration), Phys. Lett. B {\bf 723}, 82 (2013).

\bibitem{threshold2}J.~Haidenbauer, C.~Hanhart, X.~W.~Kang and U.~G.~Meibner, 
Phys. Rev. D {\bf 92}, 054032 (2015).

\bibitem{threshold4} M.~Ablikim {\it et al.} (BESIII Collaboration), Phys. Rev. Lett. {\bf 117}, 042002 (2016).

\bibitem{threshold3}J.~P.~Lees {\it et al.} (\babar ~Collaboration),
Phys.\ Rev.\ D {\bf 85}, 112009 (2012).

\bibitem{PDG}
C. Patrignani {\it et al.} (Particle Data Group), Chin. Phys. C {\bf 40}, 
100001 (2016).

\bibitem{Belle2009} K.~Inami {\it et al.} (Belle Collaboration),
Phys.\ Lett.\ B {\bf 672}, 209 (2009).

\bibitem{ALEPH1997} D.~Buskulic {\it et al.} (ALEPH Collaboration),
Z.\ Phys.\ C {\bf 74}, 263 (1997).

\bibitem{CLEO1992} M.~Artuso {\it et al.} (CLEO Collaboration),
Phys.\ Rev.\ Lett.\  {\bf 69}, 3278 (1992).

\bibitem{J_psi_VP_DM2}J. Jousset {\it et al.} (DM2 Collaboration),
Phys. Rev. D {\bf41}, 5 (1990).

\bibitem{J_psi_VP_MARK3} D. Coffman {\it et al.} (MARK-III Collaboration),
Phys. Rev. D {\bf38},  2695 (1988).

\end{thebibliography}
\end{document}